\documentclass[a4paper,11pt]{article}
\usepackage{graphicx}
\usepackage{amsmath}
\usepackage{citesort}
\usepackage{amsfonts}
\usepackage{amssymb}
\usepackage{amsmath}
\usepackage{latexsym}
\usepackage{srcltx}
\usepackage{url}
\usepackage{color}
\usepackage{ntheorem}
\usepackage{braket}

\newcommand{\coneD}{{\mathrm D}}

\newcounter{mnotecount}[section]

\renewcommand{\themnotecount}{\thesection.\arabic{mnotecount}}

\newtheorem{theorem}{Theorem}[section]

\newtheorem{Theorem}[theorem]{\sc  Theorem\rm}

\newtheorem{corollary}[theorem]{\sc  Corollary\rm}

\newtheorem{definition}[theorem]{\sc  Definition\rm}

\newtheorem{lemma}[theorem]{\sc Lemma\rm}
\newtheorem{Lemma}[theorem]{\sc Lemma\rm}

\newtheorem{remark}[theorem]{\sc Remark\rm}

\newtheorem{remarks}[theorem]{\sc Remarks\rm}

\newcommand{\ol}[1]{\overline{#1}{}}
\newcommand{\ul}[1]{\underline{#1}{}}

\newcommand{\jlcax}[1]{}
\newcommand{\eean}{\nonumber\end{eqnarray}}

\newcommand{\kk}[1]{}

\newcommand{\mcH}{{\mycal H}}

\newcommand{\eqs}[2]{\eq{#1}-\eq{#2}}

\newcommand{\beq}{\begin{equation}}

\newcommand{\FS}       
                  {F}

\newcommand{\HS} 
       {H_{\mbox{\scriptsize volume}}}

{\ptc{this should be removed in the oberwolfach version}}%

\newcommand{\eeal}[1]{\label{#1}\end{eqnarray}}
\newcommand{\bed}{\begin{deqarr}}
\newcommand{\eed}{\end{deqarr}}
\newcommand{\bedl}[1]{\begin{deqarr}\label{#1}}
\newcommand{\eedl}[2]{\arrlabel{#1}\label{#2}\end{deqarr}}

\newcommand{\mcO}{{\mycal O}}

\newcommand{\bel}[1]{\begin{equation}\label{#1}}
\newcommand{\bea}{\begin{eqnarray}}
\newcommand{\bean}{\begin{eqnarray}\nonumber}
\newcommand{\beal}[1]{\begin{eqnarray}\label{#1}}
\newcommand{\eea}{\end{eqnarray}}

\newcommand{\Eq}[1]{Equation~\eq{#1}}

\newcommand{\Eqs}[2]{Equations~\eq{#1}-\eq{#2}}

\newcommand{\qed}{\hfill $\Box$ \medskip}
\newcommand{\proof}{\noindent {\sc Proof:\ }}
\newcommand{\be}{\begin{equation}}
\newcommand{\eeq}{\end{equation}}
\newcommand{\ee}{\end{equation}}
\newcommand{\beqa}{\begin{eqnarray}}
\newcommand{\eeqa}{\end{eqnarray}}
\newcommand{\beqan}{\begin{eqnarray*}}
\newcommand{\eeqan}{\end{eqnarray*}}
\newcommand{\ba}{\begin{array}}
\newcommand{\ea}{\end{array}}

\newcommand{\Id}{\mbox{\rm Id}} 

\newcommand{\mcM}{{\mycal M}}
\newcommand{\mcD}{{\mycal D}}

\newcommand{\mnote}[1]
{\protect{\stepcounter{mnotecount}}$^{\mbox{\footnotesize
$
\bullet$\themnotecount}}$ \marginpar{
\raggedright\tiny\em
$\!\!\!\!\!\!\,\bullet$\themnotecount: #1} }

\newcommand{\warn}[1]
{\protect{\stepcounter{mnotecount}}$^{\mbox{\footnotesize
$
\bullet$\themnotecount}}$ \marginpar{
\raggedright\tiny\em
$\!\!\!\!\!\!\,\bullet$\themnotecount: {\bf Warning:} #1} }

\newcommand{\R}{\mathbb R}

\newcommand{\eq}[1]{(\ref{#1})}

\newcommand{\ptc}[1]{\mnote{{\bf ptc:}#1}}

\newcommand{\mcL}{{\mycal L}}

\newcommand{\beqar}{\begin{deqarr}}
\newcommand{\eeqar}{\end{deqarr}}

\newcommand{\beaa}{\begin{eqnarray*}}
\newcommand{\eeaa}{\end{eqnarray*}}

\DeclareFontFamily{OT1}{rsfs}{}
\DeclareFontShape{OT1}{rsfs}{m}{n}{ <-7> rsfs5 <7-10> rsfs7 <10-> rsfs10}{}
\DeclareMathAlphabet{\mycal}{OT1}{rsfs}{m}{n}

{\catcode `\@=11 \global\let\AddToReset=\@addtoreset}
\AddToReset{equation}{section}

{\catcode `\@=11 \global\let\AddToReset=\@addtoreset}
\AddToReset{figure}{section}

{\catcode `\@=11 \global\let\AddToReset=\@addtoreset}
\AddToReset{table}{section}

\begin{document}


\title{KIDs like cones%
\thanks{Preprint UWThPh-2013-8. Supported in part by the  Austrian Science Fund (FWF): P 24170-N16.}}
\author{
Piotr T. Chru\'sciel{}\thanks{Email  {Piotr.Chrusciel@univie.ac.at}, URL {
http://homepage.univie.ac.at/piotr.chrusciel}}\ \ and
Tim-Torben Paetz{}\thanks{Email  Tim-Torben.Paetz@univie.ac.at}   
 \vspace{0.5em}\\  \textit{Gravitational Physics, University of Vienna}  \\ \textit{Boltzmanngasse 5, 1090 Vienna, Austria }}
\maketitle

\vspace{-0.2em}

\begin{abstract}
We analyze vacuum Killing Initial Data on characteristic Cauchy surfaces. A general theorem on existence of Killing vectors
in the domain of dependence is proved, and some
special cases are analyzed in detail, including the case of bifurcate Killing horizons.
\end{abstract}

\noindent
\hspace{2.1em} PACS numbers: 04.20.Cv, 04.20.Ex

\tableofcontents

\section{Introduction}

Killing Initial Data (KIDs) are defined as initial data on a Cauchy surface  for  a spacetime Killing vector field. Vacuum KIDs on spacelike hypersurfaces are well understood~(see~\cite{MaertenKIDs,ChBeigKIDs} and references therein). In the spacelike case they play a significant role by  providing an obstruction to  gluing initial data sets~\cite{ChDelay,CorvinoSchoen2}.

The question of KIDs on light-cones has been recently raised in~\cite{TilquinSchucker}.
The object of this note is to analyze this, as well as KIDs on characteristic surfaces intersecting transversally.  It turns out that the situation in the light-cone case is considerably simpler than for the spacelike Cauchy problem, which explains our title.

For definiteness we assume the Einstein vacuum equations, in dimensions $n+1$, $n\ge 3$,
possibly with a cosmological constant,
\bel{20I13.1}
 R_{\mu\nu}=\lambda g_{\mu\nu}
 \;,
 \qquad
 \lambda \in \R
 \;.
\ee
Similar results can be proved for Einstein equations with matter fields satisfying well-behaved evolution equations.

\subsection{Light-cone}
 \label{ss14V13.2}

Consider the (future) light-cone $C_O$ issued from a point $O$ in an $(n+1)$-dimensional spacetime $(\mcM,g)$, $n\ge 3$; by this we mean the subset of $\mcM$ covered by future-directed null geodesics issued from $O$.
(We expect that our results remain true for $n=2$; this requires a more careful analysis some of the equations arising, which we have not attempted to carry out.)
Let $(x^\mu)=(x^0,x^i)=(x^0,r,x^A)$ be a coordinate system such that $x^0$  vanishes on $C_O$.
In the theorem that follows the initial data for the sought-for Killing vector field are provided by  a spacetime vector field $\ol Y $ which is defined on $C_O$ only.%
\footnote{Given a smooth vector field $Y^\mu$ defined in a spacetime neighborhood of a hypersurface $\{f=0\} $ we will write $\ol Y^ \mu = Y^\mu|_{f=0}$, but at this point of the discussion $\ol Y^\mu$ is simply a vector field defined along the surface $\{f=0\}$,  it being irrelevant whether or not $\ol Y^ \mu$ arises by restriction of a smooth space-time vector field. On the other hand, that last question will become a central issue in the proof of Theorem~\ref{T18I13.1new}.}
We will need to differentiate $\ol Y$ in directions tangent to $C_O$,
for this we need a covariant derivative operator  which involves only derivatives tangent to the characteristic hypersurfaces. In a coordinate system such that the hypersurface under consideration is given by the equation $ x^0 = 0 $, for the first derivatives the usual spacetime covariant derivative $\nabla_i \ol Y^\mu$ applies. However, the tensor of second spacetime covariant derivatives involves the undefined fields $\nabla_ 0 \ol Y^\mu$. To avoid this we set, on the hypersurface $\{x^0 =0\}$,%
\footnote{We use the following conventions on indices: Greek indices are for spacetime tensors and coordinates, small Latin letters shall be used
for tensors and coordinates on the light cone or the characteristic surfaces, and capital Latin letters for tensors or coordinates
in the hypersurfaces of spacetime co-dimension two foliating the characteristic surfaces.}
\beal{26III13.11}
 \coneD_{i}  \ol Y_ \mu
  &:= &
 \partial_ i   \ol Y_\mu -   \Gamma^\nu_{\mu i}  \ol Y_\nu \  \equiv \  \nabla_i \ol Y_\mu
 \;,
\\
 \coneD_{i} \coneD_j \ol Y_ \mu
 &: = &  \partial_ i \coneD_j \ol Y_\mu - \Gamma^k_{ij} \coneD_k \ol Y_\mu - \Gamma^\nu_{i\mu} \coneD_j \ol Y_\nu
 \nonumber
 \\
 & \equiv &  \partial_ i \nabla_j \ol Y_\mu - \Gamma^k_{ij} \nabla_k \ol Y_\mu - \Gamma^\nu_{i\mu} \coneD_j \ol Y_\nu
 \;,
\eeal{26III13.21}
with an obvious similar formula for $\coneD_{i} \coneD_j \ol Y^\mu $.
When the restriction  $\ol{\nabla_0 Y{}_ \mu }$ to the hypersurface $\{x^0 =0 \}$ of the $x^0$-derivative  is defined we have
$$ \coneD_{i} \coneD_j \ol Y_\mu = \nabla_i \nabla_j \ol Y_\mu +\Gamma^0_{ij}\ol{  \nabla_0 Y_\mu}
 \;.
$$
 Clearly, $ \coneD_{i} \coneD_j \ol Y_0 $ coincides with $\nabla _i \nabla_j   Y_0 |_{\{x^0=0\}}$ when $\ol Y ^\mu $ is the restriction to the hypersurface $\{x^0=0\}$ of a Killing vector field $Y^\mu$, as then $\nabla_0 Y_0 =0$. This is of key importance for our equations below.

In the adapted null coordinates of~\cite{CCM2} we have $\Gamma^0_{1i}=0$ on $\{x^0=0\}$ (see~\cite[Appendix~A]{CCM2}), so in these coordinates $\coneD _i\coneD_j $ differs from $\nabla _i \nabla_j$ only when $i,j\in\{2,\ldots,n\}$.

Our  main result is:
\begin{Theorem}
 \label{main_result}
Let $\ol Y$ be a continuous
vector field
defined along $C_O$ in a vacuum spacetime $(\mcM,g)$, smooth on $C_O\setminus \{O\}$.
There exists  a smooth  vector field $X$ satisfying the Killing equations on $D^+(C_O)$ and coinciding with $\ol Y$ on $C_O$ if and only if \underline{on $C_O$} it holds that
\bea
   \label{condition_main}
&
  \coneD_i \ol Y_j + \coneD_j \ol Y_i  =0
  \;,
   &
\\
 &
 \label{11V13.11}
       \coneD_{1} \coneD_1\overline{Y}_0   =  R_{011}{}^{\mu}\overline Y_{\mu}
   \;.
   &
\end{eqnarray}
Furthermore, \eq{11V13.11} is not needed on the closure of the set on which the divergence $\tau$ of $C_O$ is non-zero.
\end{Theorem}

While this is not necessary, the analysis of KIDs on light-cones can thus be split into two cases: The first  is concerned with the region 
sufficiently close to the tip of the cone
where the expansion   $\tau$  has no zeros. Once a spacetime with Killing field has been constructed near the vertex, the initial value problem for the remaining part of the cone can be
reduced to a characteristic initial value problem with two transversally intersecting null hypersurfaces, which will be addressed in Theorem~\ref{T18I13.2}. From this point of view the key restriction for light-cones is \eq{condition_main}.

The proof of Theorem~\ref{main_result} can be found in Section~\ref{S_proof_main_result}.
In order to prove it we will first establish some intermediate results, Theorems~\ref{T18I13.1} and \ref{T18I13.1new} below,
which require  further hypotheses. It is somewhat surprising that these additional conditions turn out to be automatically satisfied.

Equations \eq{condition_main} provide thus necessary-and-sufficient conditions for the existence, to the nearby future of $O$, of a Killing vector field. They can be viewed as the light-cone equivalent of the spacelike KID equations, keeping in mind that \eq{11V13.11} should be added when $C_O$ contains open subsets on which $\tau$ vanishes.
As made clear by the definitions,
\eq{condition_main}-\eq{11V13.11} involve  only the derivatives of $\ol Y$ in directions tangent to $C_O$.

We shall see in Section~\ref{s9IV13.1} that some of the equations \eq{condition_main}  can be integrated to determine $\ol Y$ in terms of data at $O$. Once this has been done, we are left with  the trace-free part of $  {\coneD_A \ol Y_B + \coneD_B \ol Y_A}=0 $  as the ``reduced KID equations".

It should be kept in mind that a Killing vector field satisfies an overdetermined system of second-order ODEs which can be integrated along geodesics starting from $O$, see \eq{26I13.1} below. This provides both $X$, and its restriction $\ol X$ to $C_O$, in a neighborhood of $O$, given the free data $X^\alpha|_O$ and $\nabla^{[\alpha} X^{\beta]}|_O$.
We will see in the course of the proof of Theorem~\ref{T18I13.1} how such a scheme ties-in with the statement of the theorem, cf.\ in particular Section~\ref{sec_free_data}

In Section~\ref{s9IV13.1} we give a more explicit form of the KID equations \eq{condition_main} on a cone
and discuss some special cases.

In Section~\ref{ss14VII13.1} we show how  bifurcate Killing horizons arise from totally geodesic null surfaces normal to a spacelike submanifold $S$ of co-dimension two, and how isometries of $S$ propagate to the space-time.

\subsection{Two intersecting null hypersurfaces}

Throughout, we employ the symbol ``$\ \breve{\cdot}\ $" to denote the trace-free part of the  field ``$\ \cdot\ $" with respect to $\tilde g = \ol g_{AB}\, \mathrm{d}x^A\mathrm{d}x^B$.   Further, an overbar denotes restriction to the initial surface.

In what follows the coordinates $x^A$ are assumed to be constant on the generators of the null hypersurfaces.
The analogue  of Theorem~\ref{main_result} for two intersecting hypersurfaces reads:
\begin{Theorem}
 \label{T18I13.2}
Consider two smooth null hypersurfaces $N_1=\{x^1=0\}$ and $N_2=\{x^2=0\}$ in an $(n+1)$-dimensional  vacuum spacetime $(\mcM,g)$,
with transverse intersection along a smooth $(n-1)$-dimensional submanifold $S$.
Let $\ol Y$ be a continuous vector field
defined on $N_1\cup N_2$ such that $\ol Y|_{N_1}$ and $\ol Y|_{N_2}$ are smooth. There exists  a smooth  vector field $X$ satisfying the Killing equations on $D^+(N_1
\cup N_2)$ and coinciding with $\ol Y$ on $N_1\cup N_2$  if and only if \underline{on $N_1$} it holds that
\begin{eqnarray}
&
\coneD_2 \overline{Y}_2 =0
\label{hyp_cond1}
 \;,
&
\\
&
 \coneD_{(2} \ol Y_{A)}  =0
\label{hyp_cond2}
 \;,
&
\\
&
 (\coneD_{(A} \ol Y_{B)})\breve{}  =0
\label{hyp_cond3}
 \;,
&
\\
 &
  R{}_{122}{}^{\mu} \ol Y_{\mu}-\coneD_{2}\coneD_2\ol Y_1 = 0
\label{hyp_cond4}
  \;,
\end{eqnarray}
where $\coneD$ is the analogue on  $N_1$ of the derivative operator \eq{26III13.11}-\eq{26III13.21}, with  identical corresponding conditions  on $N_2$, and on $S$ one needs further to assume that
\begin{eqnarray}
 &
 (\coneD_1 \ol Y_{2} + \coneD_{2}\ol Y_1)|_S = 0
 \;,
 &
\label{hyp_cond_S}
\\
&
g^{AB} \coneD_A \ol Y_B |_S =0
 \label{hyp_cond2a}
\;,
&
\\
&
 \partial_i (g^{AB}\coneD_A \ol Y_B )|_S =0\;, \quad i=1,2
 \label{hyp_cond2aa}
\;,
&
  \\
 &
   R_{21A}{}^{\mu} \ol Y_{\mu}  -\coneD_{A}\coneD_{[1} \ol Y_{2]} |_S = 0
   \;.
 &
    \label{hyp_cond2b}
\end{eqnarray}
\end{Theorem}

Similarly to Theorem~\ref{main_result}, \eq{hyp_cond4} can be replaced by the requirement  that $\ol g^{AB} \coneD_A \ol Y_B=0$ on regions where the divergence
of $N_1$ is non-zero. An identical statement applies to $N_2$.

Theorem~\ref{T18I13.2} is proved in Section~\ref{s18III13.2}. As before,
\eqs{hyp_cond1}{hyp_cond2b} provide necessary-and-sufficient conditions for the existence, to the future of $S$, of a Killing vector field. Hence they provide the equivalent of the spacelike KID equations in the current setting.
Note that in \eq{hyp_cond1}-\eq{hyp_cond2b}  the derivative $\coneD$ coincides with $\nabla$.

\section{The light-cone case}
 \label{s18III13.3}

\subsection{Adapted null coordinates}
 \label{ss14V13.1}

We use local coordinates $(x^0\equiv u,x^1\equiv r, x^A)$ adapted to the light-cone as in~\cite{CCM2}, in the sense that the cone is given by $C_{O}=\{x^{0}=0\}$ . Further, the coordinate $x^1$ parameterizes the null geodesics emanating to the future from the vertex of the cone, while the $x^A$'s are local coordinates
on the level sets $\{x^0=0,\,x^1=\mathrm{const}\}\cong S^{n-1} $, and are constant along the generators.
Then the metric takes the following form on $C_{O} $:
\begin{equation}
 g|_{C_O} = \overline{g}_{00} (dx^{0})^{2} + 2 \nu_0 dx^{0} dx^{1} + 2 \nu_A  dx^{0} dx^{A} +
 \overline{g}_{AB} dx^{A} dx^{B}
 \;.
\end{equation}
We stress that we \emph{do not} assume that this form of the
metric is preserved under differentiation in the
$x^{0}$--direction, i.e.\ we do not impose any gauge condition off the cone.
On $C_O$ the inverse metric reads
\begin{equation}
\label{29VIII.1}
g^{\sharp}|_{C_O} = \overline{g}^{11}\partial^{2}_{1} + 2 \overline{g}^{1A} \partial_r  \partial_{A}
+ 2 \nu^0 \partial_{0} \partial_r  + \overline{g}^{AB}\partial_{A} \partial_{B}
  \;,
\end{equation}
with
\begin{equation}
\nu^0 = \frac{1}{\nu_0}
\;,
\quad
\overline{g}^{1A} = -\nu^0 \overline{g}^{AB} \nu_B
\;,
\quad
\overline{g}^{11} = (\nu^0)^{2} (-\overline{g}_{00} + \overline{g}^{AB}\nu_A\nu_B)
\;.
\end{equation}

\subsection{A weaker result}
 \label{ss18III13.3}

We start with a weaker version of Theorem~\ref{T18I13.1new} which, moreover, assumes that the vector field $\ol Y$ there is the restriction to the light-cone of some smooth vector field $Y$:

\begin{Theorem}
 \label{T18I13.1}
Let $Y$ be a smooth vector field defined in a neighborhood of $C_O$ in a vacuum spacetime $(\mcM,g)$. There exists  a smooth  vector field $X$ satisfying the Killing equations on $D^+(C_O)$ and coinciding with $Y$ on $C_O$ if and only if the equations
\bea
   \label{20I13.2nx}
   \label{18I13.2a}
&
  \coneD_i \ol Y_j + \coneD_j \ol Y_i  =0
 \;,
&
\\
 &
   \label{20I13.2new}
   \label{20I13.2}
   R_{011}{}^{\mu}\overline Y_{\mu}-\coneD_{1}\coneD_1\overline{Y}_0 = 0
  \;,
 &
  \\
 &
   \label{20I13.2bnew}
   \label{20I13.2b}
   R_{01A}{}^{\mu}\overline Y_{\mu}  -\coneD_{A}\coneD_1 \overline{Y}_0 = 0
   \;,
 &
   \\
 &
   \label{20I13.2c}
   \label{20I13.2cnew}
    g^{AB}(  R_{0AB}{}^{\mu}\overline Y_{\mu} -  \coneD_{A} \coneD_B\overline{Y}_0  ) = 0
   \;,
   &
\end{eqnarray}
are satisfied by the restriction $\ol Y$ of $Y$ to $C_O$.
\end{Theorem}

{\noindent\sc Proof:}
To prove necessity, let $X$ be a smooth vector field satisfying the Killing  equations on $D^+(C_O)$:
\bel{20XI12.1}
\underbrace{\nabla_\mu X_\nu + \nabla_\nu X_\mu}_{=:A_{\mu\nu}} =0
\;,
\ee
the tangent components of which give   \eq{18I13.2a}.
It further easily follows from \eq{20XI12.1} that $X$  satisfies
\bel{18I13.1}
\nabla_{\mu}\nabla_{\nu} X_\sigma = R _{\alpha\mu \nu\sigma}  X^\alpha
\;,
\ee
and   \eq{20I13.2}-\eq{20I13.2c} similarly follow; for \eq{20I13.2c} the equation $\ol A_{00}=0$ is used.

To prove sufficiency, by contracting \eq{18I13.1} one finds
\bel{20XI12.2}
 \Box X^\sigma = - R^\sigma{}_\alpha X^\alpha
\ee
(which equals $ -\lambda X^\sigma$ under \eq{20I13.1}).
So, should a solution $X$ of our problem exist,  it will necessarily satisfy the wave equation \eq{20XI12.2}.

Now, it follows from e.g.~\cite[Th\'eor\`eme~2]{Dossa97}
that for any smooth vector field $Y $ defined on $\mcM$ there exists a smooth vector field $X$ on $\mcM$ solving \eq{20XI12.2} to the future of $O$, such that,
\bel{20XI12.3}
 \overline X^\mu =  \overline Y^\mu
 \;.
\ee
Here, and elsewhere, overlining denotes  restriction to $C_O$.
Further $X|_{D^+(C_O)}$ is uniquely defined by  $Y|_{C_O}$.

Applying $
\Box$ to  \eq{20XI12.1}
leads to the identity
\begin{equation}
 \Box A_{\mu\nu} = 2\nabla_{(\mu}\Box X_{\nu)} +  4R_{\kappa(\mu\nu)}{}^{\alpha}\nabla^{\kappa}X_{\alpha} + 2X_{\alpha} \nabla^{\kappa}R_{\kappa(\mu\nu)}{}^{\alpha}
  +2R_{\alpha(\mu}\nabla^{\alpha} X_{\nu)}
   \;.
   \label{18I13.7asdf-}
\end{equation}
When $X^\mu$ solves \eq{20XI12.2}  this can be rewritten as a homogeneous linear   wave equation for the tensor field $A_{\mu\nu}$,
\begin{eqnarray}
 \Box A_{\mu\nu} &= &   -2R_{\mu}{}^{\alpha}{}_{\nu}{}^{\kappa}A_{\alpha\kappa} +2R_{(\mu}{}^{\alpha}A_{\nu)\alpha} -2\mcL_X R_{\mu\nu}
   \;,
   \label{18I13.7asdf}
\end{eqnarray}
if one notes that, under \eq{20I13.1} the last term  $-2\mcL_X R_{\mu\nu}$ equals $  -2\lambda A_{\mu\nu}$ (and, in fact, cancels with the before-last one, though this cancellation is irrelevant for the current discussion).
It follows from uniqueness of solutions of \eq{18I13.7asdf} that a solution $X$ of \eq{20XI12.2} will satisfy the Killing equation on $D^+(C_O)$ if and only if
\bel{20XI12.5}
 \overline{A}{_{\mu\nu}}=0
 \;.
\ee
But by   \eq{18I13.2a} we already have
\bel{20XI12.5a}
 \overline{A}{_{ij}}=0
 \;,
\ee
so it remains to show that the equations $\overline{A}{_{0\mu}}=0$ hold. (Annoyingly, these equations involve the derivatives $\ol{\partial_0 X_\mu}$ which \emph{cannot} be expressed as \emph{local} expressions involving only the initial data $\ol X = \ol Y$.) The theorem follows now directly from Lemma~\ref{L19IV13.2} below.
\qed

\begin{definition}
 It is convenient to introduce, for a given vector field X, the tensor field
\begin{equation}
 S^{(X)}_{\mu\nu\sigma} := \nabla_{\mu}\nabla_{\nu}X_{\sigma} - R^{\alpha}{}_{\mu\nu\sigma}X_{\alpha}
\;.
 \label{definition_S}
\end{equation}
 Whenever it is clear from the context which vector field is meant we will suppress its appearance and simply write $S_{\mu\nu\sigma}$.
\end{definition}

Using the algebraic symmetries of the Riemann tensor we find:

\begin{lemma}
\label{properties_S}
It holds that:
\begin{enumerate}
 \item[(i)] $2S_{\alpha\beta\gamma} = 2 \nabla_{(\alpha}A_{\beta)\gamma} - \nabla_{\gamma}A_{\alpha\beta}$,
\item[(ii)]  $2S_{\alpha(\beta\gamma)} = \nabla_{\alpha}A_{\beta\gamma}$,
\item[(iii)] $S_{[\alpha\beta]\gamma}=0$.
\end{enumerate}
\end{lemma}

\begin{Lemma}
 \label{L19IV13.2}
 Suppose that $\ol A_{ij}=0$ and $\Box X= -\lambda X$.
 \begin{enumerate}
 \item
   \eq{20I13.2}  is equivalent to
 $
  \ol A_{01}=0
  \;.
 $
 \item If  \eq{20I13.2} holds, then   \eq{20I13.2b} is equivalent to
 $
  \ol A_{0A}=0
  \;.
 $
 \item If \eq{20I13.2} and    \eq{20I13.2b} hold, then  \eq{20I13.2cnew} is equivalent to
 $
  \ol A_{00}=0
  \;.
 $
 \end{enumerate}
\end{Lemma}

 {\noindent \sc Proof:}
It turns out to be convenient to consider the identity
\beaa
 \nabla^\mu A_{\mu\nu} &\equiv &  \underbrace{\nabla^\mu(\nabla_{\mu}X_{\nu}}_{=-R_{\nu\mu} X^\mu} +\nabla_{\nu}X_{\mu} ) =-R_{\nu\mu} X^\mu+ \nabla^\mu \nabla_{\nu}X_{\mu} = \nabla_\nu \nabla^{\mu}X_{\mu}
\\
 & = &  \frac 12 \nabla_\nu A^\mu{}_\mu
 \;.
\eeaa
Thus, it holds that
\beal{18I13.5}
  g^{\alpha\beta}(2 \nabla_\alpha A_{\beta\nu} - \nabla_\nu A _{\alpha\beta}) = 0
 \;.
\eea
In adapted null coordinates  \eq{18I13.5}  implies
\bean
 0
  & = &
  2\nu^0 ( \ol{\nabla_0   A_{1\mu}}+ \nabla_1 \ol  A_{0 \mu} -  \nabla_{\mu} \ol A _{01}) +
  2\ol g^{1B}( \nabla_1 \ol  A_{B \mu}
  + \nabla_B \ol  A_{1 \mu} - \nabla_{\mu} \ol A _{1B})
\\
 &&
  +
  \ol g^{11}(2 \nabla_1 \ol  A_{1 \mu} - \nabla_{\mu} \ol A _{11})
  +
  \ol g^{BC}(2 \nabla_B \ol  A_{C \mu} - \nabla_{\mu} \ol A _{BC})
 \;.
\eeal{18I13.6-1}
Due to Lemma~\ref{properties_S} we have
\begin{eqnarray}
 \nabla _{0}A_{i j  }
 & = &
   \nabla_i  A_{j  0}  +  \nabla_j   A_{i 0}
    - 2S_{ij0}
\label{18I13.10}
  \;.
\end{eqnarray}
When $(ij)=(11)$ that yields
\bea
 \nabla_0 A_{11} -  2\nabla_1  A_{01} =  -2S_{110}
 \;.
\eeal{18I13.7}
Inserting into \eq{18I13.6-1} with $\nu=1$, after some simplifications one obtains
\beaa
\lefteqn{
  4\nu^0{\nabla_1  \ol A_{01}}+2   \ol g^{1B} \nabla_B\ol  A_{11}
   }
   &&
\\ &&
  - \ol g^{BC}(  \nabla_1 \ol A_{BC }-2 \nabla_B\ol  A_{C1} )+ \ol g^{11}\nabla_1\ol  A_{11}
   = 4 \nu^0 \ol S_{110}
 \;.
\eeaa
Using the vanishing of the $\ol\Gamma^0_{i1}$'s~\cite[Appendix~A]{CCM2} and the $\ol A_{ij}$'s,  this becomes a linear homogeneous  ODE for $\ol A_{01}$; in the notation of the last reference (where, in particular, $\tau$ denotes the divergence of $C_O$):%
\footnote{Throughout we shall make  extensively use of the formulae for the Christoffel symbols in adapted null coordinates computed in~\cite[Appendix~A]{CCM2}.
Apart from the  vanishing of the $\ol\Gamma^0_{i1}$'s the expressions for $\ol \Gamma^0_{01}$, $\ol \Gamma^1_{11}=\kappa$, $\ol \Gamma^A_{1B}$ and
$\ol \Gamma^0_{AB}$ will be often used.}
\begin{eqnarray}
 (2 \partial_r + \tau-2\nu^0\partial_r \nu_0 )\ol A_{01} =   2\ol S_{110}
 \;.
  \label{15III13.1}
\end{eqnarray}
If $\ol A_{01}=0$, the vanishing of $\ol S_{110}$ immediately follows.

To prove the reverse implication, for definiteness  we assume here and in what follows  a coordinate system as in~\cite[Section~4.5]{CCM2}. In this coordinate system $\tau$ behaves as $(n-1)/r$ for small $r$,
$\nu_0$ satisfies $\nu_0=1+O(r^2)$,
and
\eq{15III13.1} is a Fuchsian ODE with the property  that every solution which is $o(r^{-(n-1)/2})$ for small $r$ is identically zero, see Appendix~\ref{A5VIII13.1}.
As $\ol A_{01}$ is bounded,
when $\ol S_{110}$ vanishes we conclude that
\bel{18I13.9}
 \ol A_{01}=0
 \;.
\ee
This proves point 1 of the lemma.

Next, \eq{18I13.6-1} with $\nu=D$ reads
\bean
 0
  & = &
  2\nu^0( \ol{\nabla_0   A_{1 D}}+ \nabla_1 \ol  A_{0 D} -  \nabla_D \ol A _{01})
\\
 \nonumber
  &&
     +
 2 \ol g^{1B}(  \nabla_1 \ol  A_{B D}
  +  \nabla_B \ol  A_{1D} - \nabla_D \ol A _{1B})
\\
 &&
  +
  \ol g^{11}(2 \nabla_1 \ol  A_{1 D} - \nabla_D \ol A _{11})
  +
  \ol g^{BC}(2 \nabla_B \ol  A_{C D} - \nabla_D \ol A _{BC})
 \;.
\eeal{18I13.6-2}
Using \eq{18I13.10} with $(ij)=(A1)$,
\begin{eqnarray}
 \nabla_0A_{1A} &=& \nabla_AA_{01} + \nabla_1A_{0A} -2S_{A10}
\label{eqn_trans_A_1A}
 \;,
\end{eqnarray}
 to eliminate  $ \ol{\nabla_0   A_{1 D}}$  from \eq{18I13.6-2}, and invoking   \eq{18I13.9},
on $C_O $ one obtains a system of Fuchsian radial ODEs  for $\ol A_{0D}$,
\begin{eqnarray}
 (2\partial_r  + \frac{n-3}{n-1}\tau +2\kappa -2\nu^0\partial_r \nu_0)\overline A_{0B}    -2\sigma_B{}^C\overline A_{0C}  = 2\ol  S_{B10}
\label{15III13.1b}
\;,
\end{eqnarray}
with zero being the unique solution with the required behavior at $r=0$
when $\ol S_{B10}=0$:
\bel{18I13.11}
 \ol A_{0B}=0
 \;.
\ee
This proves point 2 of the lemma.

Let us finally turn attention to \eq{18I13.6-1} with $\nu=0$:
\bean
 0
  & = &
  2 \nu^0 \nabla_1 \ol  A_{0 0}  +
  2\ol g^{1B}(  \nabla_1 \ol  A_{B 0}
  +  \nabla_B \ol  A_{1 0} -  \ol{\nabla_0 A _{1B}})
\\
 &&
  +
  \ol g^{11}(2 \nabla_1 \ol  A_{1 0} - \ol {\nabla_0 A _{11}})
  +
  \ol g^{BC}(2 \nabla_B \ol  A_{C 0} -\ol{ \nabla_0 A _{BC}})
 \;.
\eeal{18I13.6-3}
The transverse derivatives $ \ol{\nabla_0 A _{11}}$ and $ \ol{\nabla_0 A _{1A}}$  can be eliminated using \eq{18I13.7} and \eq{eqn_trans_A_1A},
\begin{eqnarray}
 2\nu^0\nabla_1\ol A_{00} + 4\ol g^{1B}\ol S_{B10} + 2\ol g^{11}\ol S_{110} + \ol g^{BC}(2\nabla_B\ol A_{0C} - \ol{\nabla_0A_{BC}}) =0
 \;.
  \phantom{xx}
\label{18I13.6-3add}
\end{eqnarray}
The remaining one,  $ \ol g^{AB}\ol{\nabla_0 A _{AB}}$, fulfills the following equation on $C_O$, which follows from \eq{18I13.10},
\begin{eqnarray*}
 \ol g^{AB} \ol{\nabla_0A_{AB}}
 &=& 2 \ol g^{AB}\nabla_A \ol A_{0B} - 2  \ol g^{AB} \ol S_{AB0}
\\
 &=& 2 \ol g^{AB}\nabla_A \ol A_{0B} -2\tilde S - \tau \nu^0 \ol A_{00}
\;,
\end{eqnarray*}
where we have set
\begin{equation}
 \tilde S := \ol g^{AB} \ol S_{AB0} -\frac{1}{2}\tau \nu^0 \ol A_{00}
 \;.
\end{equation}
Note that $\tilde S$ is the negative of the left-hand side of \eq{20I13.2cnew}, and we want to show that the vanishing of $\tilde S $ is equivalent to that of $\ol A_{00}$.
Equation \eq{18I13.6-3add} with $\overline A_{ij}=0$ and $\overline A_{0i}=0$ (i.e.\ $\ol S_{i10}=0$) yields
\begin{eqnarray}
   \nu^0 (2 \partial_r   + \tau  + 4 \kappa - 4\nu^0\partial_r \nu_0  )  \ol A_{00} = -2\tilde S
\label{15III13.1c}
 \;.
\end{eqnarray}
For $\tilde S=0$ this is again a Fuchsian radial ODE for  $ \ol  A_{0 0} $, with the only regular solution $ \ol  A_{00}=0$, and the lemma is proved.
\qed
\medskip

\subsection{The free data for $X$}
\label{sec_free_data}

Let us  explore the nature of \eq{condition_main}.
Making extensive use of~\cite[Appendix~A]{CCM2}, and of the notation there  (thus $\kappa\equiv \overline\Gamma{}^1_{11}$, $\xi_A\equiv  -2\overline \Gamma{}^{1}_{1A}$, while  $\chi_{A}{}^B=\ol \Gamma_{1A}^B$ denotes the null second fundamental form of $C_O$),
we find
\begin{eqnarray}
 \label{13III13.10}
 \overline A_{11} &=& 2(\partial_r -\kappa)\overline  X_1
  \;,
\\
 \overline A_{1A} &=& \partial_r \overline X_A - 2\chi_A{}^B\overline X_{B}  + (\partial_A + \xi_A)\overline X_{1}
  \;,
 \label{13III13.11}
\\
 \overline A_{AB} &=& 2\tilde \nabla_{(A} \overline X_{B)} + 2\chi_{AB}\overline X{}^1 - \nu^0(2\tilde\nabla_{(A}\nu_{B)}-\overline{\partial_0 g_{AB}})\overline X_1
\;.
 \label{13III13.12}
\end{eqnarray}

For definiteness, in the discussion that follows we continue to assume  a coordinate system as in~\cite[Section~4.5]{CCM2}, in particular $\kappa=0$ and
\begin{eqnarray}
 \label{7VI.40}
 &
 \chi_{A}{}^{B} = \frac{1}{r} \delta_{A}{}^{B} + O(r)\;,\quad
 \xi_A =O(r^2)\;,
 &
\\
 \label{7VI.41}
 &
 \tau = \frac{n-1}{r} + O(r)\;,\quad
 \partial_{r}(\tau - \frac{n-1}{r}) = O(1)\;,\quad
 \partial_{A} \tau = O(r) \;,
 &
\\
 \label{7VI.42}
 &
 \sigma_{A}{}^{B} = O(r)\;,\quad
 \partial_{r} \sigma_{A}{}^{B} =  O(1)\;,\quad
 \partial_{C} \sigma_{A}{}^{B} =  O(r)\;.
 &
\end{eqnarray}

Under \eq{condition_main}  the left-hand sides of \eq{13III13.10}-\eq{13III13.12} vanish.  Hence,  we can determine $\ol X_1$ by integrating \eq{13III13.10},
\bel{17III13.1x}
 \ol X_1 (r,x^A) =  c(x^A)
   \;,
\ee
for some function of the angles.

We continue by integrating \eq{13III13.11}. This is a Fuchsian ODE for $\ol X_B$, the solutions of which are of the form
\begin{eqnarray}
 \overline X{}_A &=&r \mathring\mcD_Ac  +  f_A(x^B)r^2 +O(r^3)
 \;,
  \label{17III13.2}
\end{eqnarray}
where $\mathring\mcD $ is the covariant derivative operator of the unit round metric $s$ on $S^{n-1}$ and where $f_A(x^B)$ is an integration function.

In  a neighborhood of $O$,  where $\tau$ does not vanish,  the component $\ol X ^1$ can be algebraically determined from the equation
$\overline A_A{}^A=0$, leading to
\begin{eqnarray}
 \label{18III13.1}
 \overline X{}^1 &=&-\frac{1}{n-1}\Delta_s c -\frac{r}{n-1}s^{AB} \mathring\mcD_Af_B   + O(r^2)
 \;,
\end{eqnarray}
where $\Delta_s$ is the Laplace operator of the metric $s$.

The equation $\breve{\overline A}_{AB}=0$, where $\breve{\overline A}_{AB} $ denotes the $\tilde g$-trace free part of $ {\overline A}_{AB} $,
  imposes the relations
\begin{eqnarray}
 \label{21III13.1}
 (\mathring\mcD_A\mathring\mcD_Bc)\breve{} &=& 0  
\;,
\\
  (\mathring\mcD_{(A}f_{B)})\breve{} &=& 0
\;,
 \label{21III13.2}
\end{eqnarray}
with ${}\breve{(\cdot)}$ denoting here the trace-free part with respect to the metric $s$.

We wish, now, to relate the values of $c$ and $f_A$ to the values of the vector field $\ol X$ at the vertex, under the supplementary assumption that $\ol X$ is the restriction to $C_O$ of a differentiable vector field defined in spacetime. Following~\cite{CCM2}, we denote by $y^\mu$ normal coordinates centered at $O$. Given the coordinates $y^{\mu}$ the  coordinates
$x^{\alpha}$ can be obtained by setting
\begin{equation}
 \label{1IX.1}
x^{0}=r-y^{0}, \quad
x^{1}=r, \quad
x^{A}=\mu^{A}(\frac{y^{i}}{r})
 \;,
\end{equation}
for some functions $\mu^A $ so that the  $x^{A}$'s form local coordinates
on $S^{n-1}$, and
\begin{equation}  r:=\big\{\sum_{i}(y^{i})^{2}\big\}^{\frac{1}{2}}.
\end{equation}
We   underline the components of tensor fields in the $y^\alpha$-coordinates, in particular
\bel{15III13.2}
 \underline{ {X}{}}_\alpha= \frac{\partial x^\mu}{\partial y^\alpha}X_ \mu
 \;,
 \quad
 { {X}} _\alpha= \frac{\partial y^\mu} {\partial x^\alpha}\underline X{}_\mu
  \;,
 \quad
 \underline{ {X}} ^\alpha= \frac{\partial y^\alpha}{\partial x^\mu}X^\mu
 \;,
 \quad
 { {X}} ^\alpha= \frac{\partial x^\alpha}{\partial y^\mu} \underline X^\mu
 \;.
\ee
In particular, for   vector fields such that $\ul X^\mu$ is continuous,
we obtain
\bel{15II13.3}
 \ol X{}_1(0) =  X{}_1(0) = \ul X{}_0(0) + \sum_i \ul X{}_i(0) \frac{y^i}r= -\ul X{}^0(0)
  + \sum_i \ul X{} ^i(0) \frac{y^i}r
  \;.
\ee
Thus, for such vector fields,  $\ol X{}_1(0)$ is a linear-combination of $\ell=0$ and $\ell=1$ spherical harmonics, and contains the whole information about $\ul X^\alpha (0)$. We conclude that
\bel{17III13.1y}
     c(x^A) =-\ul X{}^0(0)
  + \sum_i \ul X{} ^i(0) \frac{y^i}r
   \;.
\ee
\Eq{21III13.1} will be satisfied if and only if
 $c$ is of the form \eq{17III13.1y}, which can be seen by noting that \eq{17III13.1y} provides a family of solutions of \eq{21III13.1} with the maximal possible dimension.

 To determine $f_A$ when $\ul X^\mu$ is differentiable at the origin we
Taylor expand $\ul X$ there,
\beaa
 \ul X_\mu = \ul X_\mu (0)+ y^j \ul{\partial_j  X_\mu} (0)+ y^0 \ul{\partial_0   X_\mu }(0)+ O(|y|^2)
 \;.
\eeaa
so that
\beal{28III13.1}
  \ol X_A =  \frac{\partial y^i}{\partial x^A} \ol{  \ul X}_i
  = \frac{\partial y^i}{\partial x^A}
   \left( \ul X_i (0)+  y^j \ul{\partial_j  X_i} (0)+ y^0 \ul{\partial_0   X_i }(0)\right)+ O(r^2)
    \;,
\eea
which determines $f_A$ in terms of $\ul{\partial_\mu X_i} (0)$.
\Eq{21III13.2}, which is the conformal Killing vector field equation on $S^{n-1}$,  will be satisfied under the hypotheses of Theorem~\ref{T18I13.1} if and only if $\ul{\partial_i X_j}(0)$ is antisymmetric.

\subsection{A second intermediate result}
 \label{s11III13.1}

As a next step towards the proof of Theorem~\ref{main_result}, we drop  in Theorem~\ref{T18I13.1}  the assumption of $\ol Y$ being the restriction of a smooth spacetime vector field:

\begin{Theorem}
 \label{T18I13.1new}
Let $\ol Y$ be a  vector field defined along $C_O$ in a vacuum spacetime $(\mcM,g)$. There exists  a smooth  vector field $X$ satisfying the Killing equations on $D^+(C_O)$ and coinciding with $\ol Y$ on $C_O$ if and only if \underline{on $C_O$} it holds that
\beal{18I13.2anew}
&
  \coneD_i \ol Y_j + \coneD_j \ol Y_i  =0
 \;,
&
\\
 &
   R_{011}{}^{\mu}\overline Y_{\mu}-\coneD_{1}\coneD_1\overline{Y}_0 = 0
  \;,
 &
  \\
 &
   R_{01A}{}^{\mu}\overline Y_{\mu}  -\coneD_{A}\coneD_1 \overline{Y}_0 = 0
   \;,
 &
   \\
 &
    g^{AB}(  R_{0AB}{}^{\mu}\overline Y_{\mu} -  \coneD_{A} \coneD_B\overline{Y}_0  ) = 0
 \label{3rd_constr}
   \;.
   &
\end{eqnarray}
\end{Theorem}

{\noindent\sc Proof of Theorem~\ref{T18I13.1new}:}
 We wish to apply Theorem~\ref{T18I13.1}. The crucial step is to construct  the vector field $Y$ needed there.
For further reference
we note that \eq{3rd_constr} will not be needed for this construction.

In the argument that follows we shall ignore the distinction between $\ol X$ and $\ol Y$ whenever it does not matter.

By hypothesis it holds that
\begin{eqnarray}
\overline A_{ij}&=&0
 \label{extension_cond1}
 \;,
\\
\ol S_{i10} &=& 0
 \label{extension_cond2}
\;.
\end{eqnarray}

We define an antisymmetric tensor $\ol F_{\mu\nu}$ via
\begin{eqnarray*}
 \ol F_{ij} &:=&  \nabla_{[i}\ol X_{j]}
\;,
\\
 -\ol F_{0i } = \ol F_{i0} &:=& \nabla_i \ol X_0
\;.
\end{eqnarray*}
Then
\begin{eqnarray*}
 \overline F_{1i} &\equiv& \frac{1}{2}\nabla_1\overline{X} _{i}  - \frac{1}{2}\overline{\nabla_{i}X_1} \equiv \nabla_1 \overline{X}{_{i}} - \frac{1}{2}\overline A_{1i}
\\
 &=&  \nabla_1\overline{X}{_{i}}
 \;.
\end{eqnarray*}
Moreover,
\begin{eqnarray*}
2 \nabla_1 \ol F_{ij} &\equiv& \nabla_1\nabla_{i}\ol X_{j} - \nabla_1\nabla_{j}\ol X_{i}
\\
 &\equiv& \nabla_{i}\nabla_1\ol X_{j} - \nabla_{j}\nabla_1 \ol X_{i} + \overline R_{1ij}{}^{\alpha}\ol X_{\alpha} +\overline R_{j 1i}{}^{\alpha}\ol X_{\alpha}
\\
 &\equiv&  \nabla_{i}\ol A_{1j} - \nabla_{j}\ol A_{1i} - \ol{\nabla_i\nabla_jX_1} + \ol{\nabla_j\nabla_iX_1} - \ol R_{ij1}{}^{\alpha}\ol X_{\alpha}
\\
 &\equiv&   \nabla_{i}\ol A_{1j} - \nabla_{j}\ol A_{1i}  - 2\ol  R_{ij 1}{}^{\alpha}\ol X_{\alpha}
 \;.
\end{eqnarray*}
With \eq{extension_cond1} that gives
\begin{eqnarray*}
  \overline{\nabla_1 F_{ij}} &=&  -   \overline  R_{ij 1}{}^{\alpha}\overline X_{\alpha}
 \;.
\end{eqnarray*}
Further,
\begin{eqnarray*}
\nabla_1 \overline{F}{_{i0}} &\equiv& \nabla_1\nabla_i\ol  X_0 \,=\,  \nabla_i\nabla_1\ol X_0  + \ol R_{1i0}{}^{\alpha}\ol X_{\alpha}
\,=\, \ol  R_{01i}{}^{\mu}\ol X_{\mu}    + \ol R_{1i0}{}^{\alpha}\ol X_{\alpha}
\\
 &=& -\ol R_{i01}{}^{\mu}\ol X_{\mu}
  \;.
\end{eqnarray*}
To sum it up, \eq{extension_cond1} and  \eq{extension_cond2} imply that  the equations
\begin{eqnarray}
  \nabla_1 \overline{X}{_{\mu}} &=&  \overline F_{1\mu}
 \label{extension_eqn1}
\;,
\\
 \nabla_1 \overline{ F}{_{\mu\nu}}  &=&     \ol R^{\alpha}{}_{1\mu\nu }\overline X_{\alpha}
 \label{extension_eqn2}
\end{eqnarray}
hold on $C_O$,

Let  $\mathring X^\mu = \overline X^\mu|_ O$ and  $\mathring F_{\mu\nu}= \ol F_{\mu\nu}|_ O$ be the initial data at $O$ needed for solving those equations.
These  data can be calculated as follows: \eq{21III13.1} and \eq{21III13.2} show that $\mathring\mcD_A c$ and $f_A$ are conformal Killing fields on the standard sphere $(S^{n-1},s)$.
It follows from~\cite[Proposition~3.2]{Semmelmann} (a detailed exposition can be found in~\cite[Proposition~2.5.1]{SemmelmannHab}) that $c$ is a linear combination of the first two spherical harmonics, so that $\mathring X^\mu$ can be read off from $c$ using \eq{17III13.1y}.
Similarly \eq{28III13.1} can be used to read-off $\mathring F_{\mu\nu}$ from  $f_A$.

We conclude that, in coordinates adapted to $C_O$ as in \eq{1IX.1},  under the hypotheses of  Theorem~\ref{T18I13.1new} the desired Killing vector $X$ is a solution of the following problem:
\bel{26I13.1}
 \left\{
   \begin{array}{ll}
     \nabla_1 \ol{X}_\mu= \ol{F}_{1\mu} , & \hbox{on $C_O$;} \\
     \nabla_1 \ol{F}_{\alpha\beta}= R_{ \gamma 1\alpha\beta} \ol X^\gamma, & \hbox{on $C_O$;} \\
     \ol X^\mu= \mathring X^\mu, & \hbox{at $O$;} \\
     \ol F_{\mu\nu}= \mathring F_{\mu\nu}, & \hbox{at $O$;} \\
     \Box X^\mu = -\lambda X^\mu, & \hbox{on $D^+(C_O)$;}\\
     {X^\mu}= \ol{X}^{\mu}, & \hbox{on $C_O$.}
   \end{array}
 \right.
\ee

Note that  the first four equations above determine  uniquely the initial data $\overline{X}{^\mu}$ on  $C_O$ needed to obtain a unique solution of the wave equation for $X^\mu$.

Now, we claim that there exists a smooth vector field  $Y^\mu$ defined near $O$ so that $\overline X^\mu$ is the restriction of $Y^\mu$ to the light-cone. To see this, let $\mathring \ell^\mu$ be given and define $(x^\mu(s), Z^\mu(s), F_{\alpha\beta}(s))$ as the unique solution of the problem
\bel{26I13.11}
 \left\{
   \begin{array}{ll}
    \displaystyle
 \frac{d^2 x^\mu}{ds^2} + \Gamma^\mu_{\alpha\beta} \frac{dx^\alpha}{ds} \frac{dx^\beta}{ds} =0, & \hbox{} \\
    \displaystyle
    \frac{dZ_\mu}{ds}  - \Gamma^\alpha_{\mu\beta}  Z_\alpha \frac{dx^\beta}{ds}= {F_{\alpha\mu}}\frac{dx^\alpha}{ds}, & \hbox{ }
\\
    \displaystyle
\frac{d  F_{\alpha\beta}}{ds} - \Gamma^\mu_{\alpha\gamma} \frac{dx^\gamma}{ds} F_{\mu\beta}
- \Gamma^\mu_{\beta\gamma} \frac{dx^\gamma}{ds} F_{\mu\alpha} = R_{ \gamma \delta \alpha\beta} Z^\gamma \frac{dx^\delta}{ds}, & \hbox{ }  \\
    \displaystyle
     x^{\mu}(0)= 0 ,
\quad   \frac{d x^{\mu}}{ds}(0)= \mathring \ell^\mu ,&   \\
    \displaystyle
     F_{\mu\nu}(0)= \mathring F_{\mu\nu}, &   \\
    \displaystyle
     {Z^\mu}(0)= \mathring{X}^{\mu}. &
       \end{array}
 \right.
\ee
For initial values such that $x^\mu(1)$ is defined, set
\bel{26I13.3}
   \displaystyle
     Y^\mu|_{x^\mu(1)} = Z^\mu|_{s=1}.
\ee
It follows from smooth dependence of solutions of ODEs upon initial data that $Y^\mu$ is smooth in all initial variables, in particular in $\mathring \ell^\mu$. If the $x^\mu$'s are normal coordinates centered at $O$, then $x^\mu(s=1)=\mathring \ell^\mu$, which implies that \eq{26I13.3} defines a smooth vector field in a neighborhood of $O$. It then easily follows that the restriction of $Y^\mu$ to $C_O$ equals $\ol X^\mu$, as defined by the first four equations in \eq{26I13.1}.

The hypotheses of Theorem~\ref{T18I13.1} are now satisfied, and Theorem~\ref{T18I13.1new} is proved.
\qed

\subsection{Proof of Theorem~\ref{main_result}}
\label{S_proof_main_result}

To prove Theorem~\ref{main_result}
we will use Theorem~\ref{T18I13.1}, together with some of the ideas of the proof of Theorem~\ref{T18I13.1new}.
We need to show that  \eq{condition_main} together with the Einstein equations imply both the existence of a smooth extension $Y$ of $\ol Y$, and  that \eq{20I13.2}-\eq{20I13.2cnew} hold.

\subsubsection{Properties of $S_{\mu\nu\sigma}$}

Recall the definition
\begin{equation}
 S_{\mu\nu\sigma} \equiv  \nabla_{\mu}\nabla_{\nu}X_{\sigma} - R^{\alpha}{}_{\mu\nu\sigma}X_{\alpha}
\;,
\label{20IV13.1}
\end{equation}
and Lemma~\ref{properties_S}.

In the context of Theorem~\ref{main_result}, only those components of the tensor field $S_{\alpha\beta\gamma}$
which \emph{do not} involve $\partial_0$-derivatives of $X$ are a-priori known. One easily checks:

\begin{Lemma} The  components
\bel{13V13.11}
 \mbox{$\ol S_{ij\mu}$ with $ij\ne AB$}
\ee
of the restriction  to $C_O$ of   $S_{\mu\nu\sigma}$
can be algebraically determined in terms of $\ol X_\sigma\equiv \ol Y_\sigma$, $D_i \ol X_\sigma\equiv D_i \ol Y_\sigma$ and
$D_i D_j \ol X_\sigma \equiv D_i D_j \ol Y_\sigma$.
\end{Lemma}

We wish, next, to calculate $\nabla^\alpha S_{\alpha\beta\gamma}$ and $\nabla^\gamma S_{\alpha\beta\gamma}$.
This requires the knowledge of $\nabla_0 X_\mu$, of $\nabla_0 \nabla_0 X_\mu$,  and even of  $\nabla_0\nabla_0 \nabla_0 X_\mu$ in some equations.
For this, let $X$ be any extension of $\ol X$  from the light-cone to a punctured neighborhood $\mcO \setminus \{O\}$ of $O$, so that the transverse derivatives appearing in the following equations
are defined. $X$ is assumed to be smooth on its domain of definition, and we emphasise that we do not make any hypotheses on the behavior of the extension $X$ as the tip $O$ of the light-cone is approached.
As will be seen, the transverse derivatives of $X$ on $C_O$   drop out from those final formulae which are relevant for us.

We will make use several times of
$$
 \nabla_\alpha R^\alpha{}_{\beta\gamma\delta} =0
 \;,
$$
which is a standard consequence of the second Bianchi identity when  the Ricci tensor is proportional to the metric.

We start with  $\nabla^\alpha S_{\alpha\beta\gamma}$. Two commutations of derivatives allow us to rewrite the first term in the divergence of $ S_{\alpha\beta\gamma}$ over the first index as
\bean
  \nabla^\alpha  \nabla_\alpha\nabla_\beta X_\gamma
   & = &
  \nabla^\alpha ( \nabla_\beta \nabla_\alpha X_\gamma + R_\gamma{}^\sigma{}_{\alpha\beta} X_ \sigma)
\\
   & = &
 \nabla_\beta \nabla^\alpha \nabla_\alpha X_\gamma
 + R {}^{\alpha\sigma}{}_{\alpha\beta} \nabla_ \sigma X_ \gamma
 + 2R_\gamma{}^\sigma{}_{\alpha\beta} \nabla^\alpha X_ \sigma
  \nonumber
\\
 & = &
 \nabla_\beta \Box X_\gamma
 + R {}^{ \sigma}{}_{ \beta} \nabla_ \sigma X_ \gamma
 + 2R_\gamma{}^\sigma{}_{\alpha\beta} \nabla^\alpha X_ \sigma
 \;.
\eeal{20IV13.2}
Hence, since $R_{\alpha\beta}=\lambda g_{\alpha\beta}$, and using the first Bianchi identity in the second line
\bean
  \nabla^\alpha  S_{\alpha \beta \gamma}
   & = &
 \nabla_\beta \Box X_\gamma
 + \lambda  \nabla_ \beta X_ \gamma
 + 2R_\gamma{}^\sigma{}_{\alpha\beta} \nabla^\alpha X_ \sigma
 - R^\sigma{}_{\alpha\beta\gamma{}} \nabla^\alpha X_ \sigma
\\
   & = &
 \nabla_\beta \left( \Box X_\gamma
 + \lambda   X_ \gamma \right)
 +  R_\gamma{}^{\sigma\alpha}{}_{\beta} A_{\alpha \sigma}
 \;.
\eeal{20IV13.3}
Similarly,
\bean
 \nabla^\alpha S_{\beta\gamma\alpha}
  & = & \nabla^\alpha \left(\nabla_\beta \nabla_\gamma X_\alpha - R^\sigma {}_{\beta\gamma\alpha} X_\sigma\right)
\\
 \nonumber
  & = & \nabla_\beta \nabla^\alpha \nabla_\gamma X_\alpha
   + R_{\gamma}{}^{\sigma\alpha}{}_{\beta}    \nabla_\sigma X_\alpha
   + R_{\alpha}{}^{\sigma\alpha}{}_{\beta}    \nabla_\gamma X_\sigma
   - R^\sigma {}_{\beta\gamma\alpha} \nabla^\alpha  X_\sigma
\\
 \nonumber
  & = & \nabla_\beta \left( \nabla_\gamma\nabla^\alpha X_\alpha
   + R {}^{\sigma }{}_{\gamma} X_\sigma
   \right)
   + R_{\gamma}{}^{\sigma\alpha}{}_{\beta}    \nabla_\sigma X_\alpha
+ R {}^{\sigma }{}_{\beta}    \nabla_\gamma X_\sigma
\nonumber
\\
  &&    - R^\sigma {}_{\beta\gamma\alpha} \nabla^\alpha  X_\sigma
   \nonumber
\\
  & = & \frac{1}{2}\nabla_\beta  \nabla_\gamma A^\alpha{}_\alpha
   + \lambda A_{\beta \gamma}
 \;.
\eeal{20IV13.4}
Now, on $C_O$ and in coordinates adapted to the cone
\bean
  \ol{\nabla^\alpha  S_{\alpha \beta \gamma}}
    & = &
      \nu^0 \left(\ol{\nabla_0 S_ { 1\beta\gamma} }+  \nabla_1 \ol S_ {0 \beta\gamma}\right)
      +
      \ol g^{1A} \left(\nabla_1\ol  S_ { A\beta\gamma} +  \nabla_A\ol  S_ {1 \beta\gamma}\right)
\\
 &&
      +
      \ol g^{11} \nabla_1 \ol S_  {1 \beta\gamma}
      +
      \ol g^{AB} \nabla_A \ol S_ {B \beta\gamma}
 \;,
\eeal{20IV13.5}
while
\bean
 \ol { \nabla^\alpha  S_{  \beta\gamma\alpha}}
    & = &
      \nu^0\left(\ol {\nabla_0 S_ { \beta\gamma 1}} +  \nabla_1 \ol S_ { \beta\gamma  0}\right)
      +
      \ol g^{1B} \left(\nabla_1 \ol S_ { \beta\gamma  B} +  \nabla_B \ol S_ { \beta \gamma  1}\right)
\\
 &&
      +
      \ol g^{11} \nabla_1 \ol S_  {  \beta\gamma  1}
      +
      \ol g^{BC} \nabla_B \ol S_ { \beta \gamma  C}
 \;.
\eeal{20IV13.6}
In order to handle undesirable terms such as $\nabla_0 S_{ \beta\gamma 1}$ we write
\bean
  \nabla_0  \nabla_\alpha\nabla_\beta X_\gamma
   & = &
 \nabla_\alpha \nabla_0\nabla_\beta X_\gamma
   + R_{\beta}{}^\sigma{}_{0 \alpha} \nabla_{\sigma} X_\gamma
  + R_\gamma{}^\sigma{}_{0 \alpha} \nabla_{\beta} X_ \sigma
\\
   & = &
 \nabla_\alpha \left(\nabla_\beta \nabla_0 X_\gamma +
  R_ {\gamma}{}^{\sigma}{}_{0 \beta}   X_ \sigma\right)
   + R_{\beta}{}^\sigma{}_{0 \alpha} \nabla_{\sigma} X_\gamma
  + R_\gamma{}^\sigma{}_{0 \alpha} \nabla_{\beta} X_ \sigma
  \nonumber
\\
   & = &
 \nabla_\alpha \left(\nabla_\beta (A_{0\gamma}- \nabla_\gamma  X_0) +
  R_ {\gamma}{}^{\sigma}{}_{0 \beta}   X_ \sigma\right)
  \nonumber
\\
 &&  + R_{\beta}{}^\sigma{}_{0 \alpha} \nabla_{\sigma} X_\gamma
  + R_\gamma{}^\sigma{}_{0 \alpha} \nabla_{\beta} X_ \sigma
 \;.
\eeal{22IV13.1}

\subsubsection{Analysis of condition \eq{20I13.2}}

\begin{lemma}
\label{lem_relation_trA_S110}
Assume that $\ol A_{1i}=0$ and $\breve{\ol A}_{AB}=0$. Then, in vacuum,
\begin{equation}
  (\partial_r  + \frac{2}{n-1}\tau -\kappa)\partial_r  (\ol g^{AB} \ol A_{AB})
 =2 \tau \nu^0 \ol S_{110}
 \;.
 \label{relation_traceA_S110}
\end{equation}
\end{lemma}

{\noindent\sc Proof:}
By Lemma~\ref{properties_S}  the vanishing of  $\ol A_{1i}$ implies
\begin{enumerate}
 \item[(i)]  $\ol S_{111} =0$,
\item[(ii)] $0=\ol S_{11A} $, as well as  all permutations thereof.
\end{enumerate}

Consider \eq{20IV13.4} with $(\beta\gamma)=(11)$. Setting $a:= \ol g^{AB}\ol A_{AB}$
we find
\begin{eqnarray}
 \nonumber
 \ol{\nabla^{\alpha}S_{11\alpha}} &=& \frac{1}{2}\nabla_1\nabla_1\ol A_{\alpha}{}^{\alpha} + \lambda \ol A_{11}
 =  \nu^0\nabla_1\nabla_1\ol A_{01}
  + \frac{1}{2}\ol g^{AB} \nabla_1\nabla_1 \ol A_{AB}
\\
 &=& \nu^0\nabla_1\nabla_1\ol A_{01}
  + \frac{1}{2}(\partial_r -\kappa)\partial_r  a
\;.
\label{13V13.12}
\end{eqnarray}
Due to Lemma~\ref{properties_S} we have
\begin{eqnarray*}
 2\chi^{AB}\ol S_{1AB} &=& \chi^{AB}\nabla_1  \ol A_{AB}
= \chi^{AB}(\partial_r \ol A_{AB} - 2 \chi_A{}^C\ol A_{BC})
\\
&=& \frac{1}{n-1}\chi^{AB}[\partial_r (a\ol g_{AB}) -2a\chi_{AB}]
\\
 &=&\frac{1}{n-1}\tau \partial_r  a
 \;.
\end{eqnarray*}
Using \eq{20IV13.6} with $(\beta\gamma)=(11)$, as well as the last equation, and employing again Lemma~\ref{properties_S} we obtain, on $C_O$,
\begin{eqnarray}
 \nonumber
 \ol {\nabla^{\alpha}S_{11\alpha}} &=& \nu^0(\ol {\nabla_0S_{111}} + \nabla_1\ol S_{110}) + \ol g^{1B}(\nabla_1\ol S_{11B} + \nabla_B\ol S_{111})
\\
 \nonumber
 && + \ol g^{11}\nabla_1\ol S_{111} + \ol g^{BC}\nabla_B\ol S_{11C}
\\
 \nonumber
 &=& \nu^0(\ol {\nabla_0S_{111}} + \nabla_1\ol S_{110}) + \ol g^{BC}\nabla_B\ol S_{11C}
\\
 \nonumber
 &=& \nu^0(\ol {\nabla_0S_{111}} + \nabla_1\ol S_{110}) - 2\chi^{AB}  \ol S_{1AB}
+\tau \nu^0\ol  S_{110}
\\
 &=& \nu^0(\ol {\nabla_0S_{111}} + \nabla_1\ol S_{110})- \frac{1}{n-1}\tau \partial_r  a + \tau \nu^0\ol  S_{110}
\;.
\label{13V13.13}
\end{eqnarray}
Using \eq{22IV13.1} we find
\begin{eqnarray}
 \nonumber
 \ol{\nabla_0 S_{111}} &=& \ol{\nabla_0\nabla_1\nabla_1X_1}=\nabla_1\nabla_1\ol A_{01} -\nabla_1 \ol S_{110} + \ol R_{011}{}^{\mu}\ol A_{1\mu}
\\
 &=& \nabla_1\nabla_1\ol A_{01} -\nabla_1 \ol S_{110}
\;.
\end{eqnarray}
Equating \eq{13V13.12} with \eq{13V13.13}, and using the last equation we end up with \eq{relation_traceA_S110}.
\qed

\begin{corollary}
\label{S110_vanishes}
 In a region where the divergence $\tau$ does not vanish  (in particular, near the vertex), $\ol A_{ij}=0$ implies, in vacuum, $\ol S_{110}=0$.
\end{corollary}

\subsubsection{Analysis of condition \eq{20I13.2b}}

\begin{Lemma}
 \label{L19IV13.1}
\label{SA10_vanishes}
Assume that  $\ol A_{ij}=0$ and $\ol S_{110}=0$. Then, in vacuum, $\ol S_{A10}=0$.
\end{Lemma}

{\noindent\sc Proof:}
From \eq{22IV13.1} we obtain
\beaa
  \nabla_0  S_{A11} & = & \nabla_0\nabla_A\nabla_1 X_1
\\  & = &
 \nabla_A  \nabla_1  A_{01}-  \nabla_A S_{110}
    + R_{1}{}^\sigma{}_{0 A} A_{\sigma 1}
 \;.
\eeaa
This allows us to rewrite \eq{20IV13.6} with $(\beta\gamma)=(A1)$ on $C_O$ as
\bean
  \lefteqn{
  \ol {\nabla^\alpha  S_{ A1  \alpha}}
    =
      \nu^0 \left(
 \nabla_A  \nabla_1 \ol  A_{01}-  \nabla_A\ol  S_{110}
    + \ol R_{1}{}^\sigma{}_{0 A} \ol A_{\sigma 1}   +  \nabla_1 \ol S_ {A1 0}\right)
    }&&
\\
 &&
      +
      \ol  g^{1B} \left(\nabla_1 \ol S_ {A1 B} +  \nabla_B\ol  S_ { A1 1}\right)
      +
      \ol g^{11} \nabla_1\ol  S_  {A1 1}
      +
    \ol   g^{BC} \nabla_B \ol S_ { A1 C}
 \;.
  \phantom{x}
\eeal{22IV13.3}
Combining with \eq{20IV13.4}, which reads with $(\beta\gamma)=(A1)$
\begin{equation*}
 \nabla^{\alpha}S_{A1\alpha} = \frac{1}{2}\nabla_A\nabla_1  A^{\alpha}{}_{\alpha} + \lambda A_{1A}\;,
\end{equation*}
 we obtain on the initial surface
\bean
   -  \nu^0 \nabla_1 \ol S_ {A1 0}
    &=&
      \nu^0 \left(
\nabla_A  \nabla_1 \ol  A_{01}- \nabla_A \ol S_{110}
    + \ol R_{1}{}^j{}_{0 A}\ol  A_{j 1}\right)
\\
 \nonumber
 &&
      +
      \ol g^{1B} \big(\nabla_1 \ol S_ {A1 B} +\nabla_B \ol S_ { A1 1}\big)
      +
      \ol g^{11} \nabla_1\ol  S_  {A1 1}
\\
  &   &
      +
     \ol  g^{BC} \nabla_B \ol S_ { A1 C}
     -\frac{1}{2}\nabla_A  \nabla_1 \ol A^\alpha{}_\alpha
   - \lambda\ol  A_{1A}
 \;.
\eeal{22IV13.4}
Assuming $\ol A_{ij}=0$,
Lemma~\ref{properties_S}  shows that  $\ol S_{i11}=0= \ol S_{1i1}=\ol S_{11i}= \ol S_{A1B}$.
This allows us to rewrite the right-hand side of \eq{22IV13.4} as
\begin{eqnarray*}
 \mbox{r.h.s.} &=&  \nu^0\nabla_A  \nabla_1  \ol A_{01} - \nu^0\nabla_A \ol S_{110} + \ol g^{BC} \nabla_B \ol S_ { A1 C}
 \\
&&  +
      \ol g^{1B} (\nabla_1 \ol S_ {A1 B} +  \nabla_B\ol S_ { A1 1}) - \frac{1}{2}\nabla_A  \nabla_1 \ol A^\alpha{}_\alpha
 \\
&=&  - \nu^0\nabla_A \ol S_{110} + \ol g^{BC} \nabla_B \ol S_ { A1 C}
  - \frac{1}{2}\ol g^{BC}\nabla_A  \nabla_1\ol A_{BC}
\\
  &&   - \ol g^{1B}(\nabla_B\ol S_{A11} - \nabla_A\nabla_1\ol A_{1B})
\;.
\end{eqnarray*}
Now, using in addition that $\ol S_{110}=0$,
\begin{eqnarray*}
 \nabla_B\ol S_{A11} - \nabla_A\nabla_1\ol A_{1B} &=& \frac{1}{2}\ol {\nabla_A (\nabla_BA_{11} - 2\nabla_1A_{1B})}
\\
 &=& -\nabla_A \ol S_{11B} \,=\,    - \nu^0 \chi_{AB} \ol S_{110} \,=\, 0
 \;.
\end{eqnarray*}
Hence,
\begin{eqnarray*}
  \mbox{r.h.s.}
 &=&  - \nu^0\nabla_A\ol  S_{110} + \ol g^{BC} \nabla_B\ol  S_ { A1 C}
  - \frac{1}{2}\ol g^{BC}\nabla_A  \nabla_1\ol A_{BC}
      \\
 &=&  - \nu^0\nabla_A \ol S_{110} + \ol g^{BC}(\nabla_B\ol S_{A1C} - \nabla_A\ol S_{B1C})
\\
 &=&  - \nu^0\nabla_A \ol S_{110} - \ol g^{BC}(\underbrace{2\chi_{[B}{}^D\ol S_{A]DC} }_{=0 \text{ by Lemma~\ref{properties_S}}}-\nu^0\chi_{BC}\ol S_{1A0} + \nu^0 \chi_{AC}\ol S_{1B0})
       \\
 &=&  - \nu^0\nabla_A\ol  S_{110}
 + \tau\nu^0 \ol S_{1A0}
 - \nu^0\chi_A{}^B \ol S_{1B0}
        \\
 &=&   \tau\nu^0 \ol S_{A10}
 + \nu^0\chi_A{}^B \ol S_{B10}
 \;,
\end{eqnarray*}
and thus, again due to $\ol S_{110}=0$ and  Lemma~\ref{properties_S},
\begin{eqnarray}
(\partial_r +\tau -\nu^0\partial_r \nu_0)\ol S_{A10}
 &=& 0
  \;.
\label{ODE_SA10}
 \end{eqnarray}
But zero is the only solution of this equation which is $o(r^{-(n-1)})$, and to be able to conclude that
\begin{eqnarray}
 \label{11V13.1}
\ol S_{A10} &=& 0
\end{eqnarray}
we need to check the behavior of $\ol  S_{A10}$ at the vertex. For definiteness we assume a coordinate system as in~\cite[Section~4.5]{CCM2}.
Now, by definition,
\beaa
\ol S_{A10} & = & \ol{\nabla_A \nabla_1 X_0 - R_{\mu A 1 0} X^\mu}
\\
 & = & \ol{\partial_A (\partial_r  X_0 - \Gamma _{10}^\mu X_\mu) - \Gamma_{A1}^\mu \nabla_\mu X_0 -
 \Gamma_{A0}^\mu \nabla_1 X_\mu- R_{\mu A 1 0} X^\mu}
 \;.
\eeaa
From \eq{17III13.1x}-\eq{18III13.1} we find
\beal{11V13.4a}
 &
 \ol X_1\;,   \partial_i \ol X_1     = O(1)
 \;,
 \quad
 \ol X_0\;,   \partial_i \ol X_0\;,  \partial_A \partial_r  \ol X_0 = O(1)
 \;,
 &
 \\
 &
 \ol X_A\;,  \partial_B \ol X_A =O(r)
 \;,
 \
 \partial_r \ol X_A = O(1)
 \;.
 &
\eeal{11V13.4b}
Using the formulae from~\cite[Appendix~A]{CCM2} one obtains
\beaa
\ol S_{A10} & = & -\ol{  \underbrace{\Gamma_{A1}^0 }_0\nabla_0 X_0  } +O(r^{-1})=O(r^{-1})
 \;,
\eeaa
which implies that \eq{11V13.1} holds, and Lemma~\ref{L19IV13.1} is proved.
\qed

\subsubsection{Proof of Theorem~\ref{main_result}}

We are ready now to prove our main result:

\medskip

{\noindent \sc Proof of Theorem~\ref{main_result}:}
By assumption, using obvious notation, $ A^{(\ol Y)}_{ij}=0$.
 When $\tau$ does not vanish
 Corollary~\ref{S110_vanishes} applies and shows that  $S^{(\ol Y)}_{110}=0$. Otherwise, $S^{(\ol Y)}_{110}=0$ holds by hypothesis and Lemma~\ref{SA10_vanishes} shows that  $S^{(\ol Y)}_{A10}$ vanishes as well.
In the proof of Theorem~\ref{T18I13.1new} we have shown that  $A^{(\ol Y)}_{ij}=0$ and $S^{(\ol Y)}_{i10}=0$ suffice to make sure that $\ol Y$
is the restriction to $C_O$ of a smooth spacetime vector field $Y$.
Then, due to the Cagnac-Dossa theorem~\cite[Th\'eor\`eme~2]{Dossa97}, there exists a smooth vector field $X$ with $\ol X=\ol Y$ which solves $\Box X=-\lambda X$.
The assertions of Theorem~\ref{main_result} follow now from Theorem~\ref{T18I13.1}, whose remaining hypotheses  are satisfied by Lemma~\ref{gAB_SAB0_vanishes}
below.
\qed

\subsubsection{Analysis of condition \eq{20I13.2cnew}}

A straightforward application of Lemma~\ref{properties_S} yields
\begin{lemma}
 \label{vanishing_S_A}
Assume that $\ol A_{i \mu}=0$. Then
\begin{enumerate}
 \item[(i)] $\ol S_{ijk}=0$,
\item[(ii)] $\ol S_{110}=\ol S_{101} = \ol S_{011} = 0$,
\item[(iii)] $\ol S_{A10} =\ol S_{A01}= \ol S_{1A0} = \ol S_{0A1} = \ol S_{10A} = \ol S_{01A} =  0$.
\end{enumerate}
\end{lemma}

\begin{lemma}
\label{gAB_SAB0_vanishes}
 Consider a smooth vector field $X$ in a vacuum spacetime $(\mcM,g)$
which satisfies  $\ol A_{i \mu}=0$ on $C_O$ and $\Box X + \lambda X=0$.
Then
\begin{equation}
 \tilde S := \ol g^{AB}\ol S_{AB0} - \frac{1}{2}\tau\nu^0 \ol A_{00}= \ol g^{AB}(\coneD_A\coneD_B \ol X_0 - \ol R_{0AB}{}^{\mu}\ol X_{\mu})=0
\;.
\end{equation}
\end{lemma}

{\noindent\sc Proof:}
Equation \eq{20IV13.3} yields with $\ol A_{i \mu}=0$, $\Box X + \lambda X=0$ and in vacuum
\begin{eqnarray*}
 \ol{g^{AB} \nabla^{\alpha}S_{\alpha AB}} \,=\, -(\nu^0)^2 \ol R_{11}\ol A_{00}
  \,=\, -\lambda (\nu^0)^2 \underbrace{\ol g_{11}}_{=0}\ol A_{00} \,=\, 0
 \;.
\end{eqnarray*}
On the other hand, \eq{20IV13.5}  gives with Lemma~\ref{properties_S} and \ref{vanishing_S_A} on $C_O$,
\begin{eqnarray*}
 \ol  g^{AB} \ol{\nabla^{\alpha}S_{\alpha AB}} &=& \nu^0 \ol g^{AB}(\ol {\nabla_0S_{1AB}} + \nabla_1\ol S_{0AB}) + \ol g^{1C}\ol g^{AB}(\nabla_1\ol S_{CAB} + \nabla_C\ol S_{1AB})
\\
 &&+ \ol g^{11}\ol g^{AB}\nabla_1\ol S_{1AB} + \ol g^{AB}\ol g^{CD}\nabla_C\ol S_{DAB}
\\
&=& \nu^0 \ol g^{AB}(\ol {\nabla_0S_{1AB}} + \nabla_1\ol S_{0AB})  + \ol g^{AB}\ol g^{CD}\nabla_C\ol S_{DAB}
\\
&=& \nu^0\ol g^{AB}(\ol {\nabla_0S_{1AB}} + \nabla_1\ol S_{0AB})  +\nu^0  \ol g^{AB}(\tau \ol S_{0AB}+ 2\chi_{A}{}^D \ol S_{D(0B)})
\\
&=&  \nu^0 \ol g^{AB}(\ol {\nabla_0S_{1AB}} - \nabla_1\ol S_{AB0} + \nabla_1 \nabla_A\ol A_{0B})
\\
&& -\tau \nu^0  \ol g^{AB} \ol S_{AB0}  + (\nu^0)^2(\tau^2 +  |\chi|^2 )\ol A_{00}
 \;.
\end{eqnarray*}
Moreover, from \eq{22IV13.1} and Lemma~\ref{properties_S}  we deduce that, on $C_O$,
\begin{eqnarray*}
\ol g^{AB}\ol {\nabla_0 S_{1AB}}
 &=&\ol   g^{AB}\ol {\nabla_0\nabla_1\nabla_AX_B}
\\
  &=&\ol  g^{AB}\ol {\nabla_1\nabla_A A_{0B}} - \ol g^{AB}\nabla_1\ol S_{AB0} + \ol g^{AB}\ol R_{01A}{}^{\mu}\ol A_{B\mu}
\\
  &=&\ol  g^{AB}\ol {\nabla_1\nabla_AA_{0B}}  - \ol g^{AB}\nabla_1\ol S_{AB0}
 \;.
\end{eqnarray*}
Hence,
\begin{eqnarray*}
 0 &=& 2 \ol g^{AB} \nabla_1 \nabla_A\ol A_{0B} - 2 \ol g^{AB} \nabla_1\ol S_{AB0}
  -\tau \ol g^{AB} \ol S_{AB0}  + \nu^0(\tau^2 +  |\chi|^2 )\ol A_{00}
\\
&=& 2 \ol g^{AB} \nabla_A \nabla_1\ol A_{0B}  -2\nu^0\ol  R_{11}\ol A_{00}   - 2 \ol g^{AB} \nabla_1\ol S_{AB0}
  -\tau \ol g^{AB} \ol S_{AB0}
\\
&&+ \nu^0(\tau^2 +  |\chi|^2 )\ol A_{00}
\\
&=& - 2 \ol g^{AB} \nabla_1\ol S_{AB0}
  -\tau \ol g^{AB} \ol S_{AB0}  + \nu^0(2\tau\nabla_1 + \tau^2 -  |\chi|^2 )\ol A_{00}
  \\
&=&  \tau \nu^0(\partial_r   + \frac{1}{2}\tau  -2\ol \Gamma^0_{01} )\ol A_{00}
 -\nu^0( \partial_r \tau - \kappa\tau + |\chi|^2) \ol A_{00}
 - 2  (\partial_r + \frac{1}{2}\tau-\ol \Gamma^{0}_{10})\tilde S
 \;.
\end{eqnarray*}
Using the vacuum constraint~\cite{CCM2} $0 = \lambda\ol g_{11} = \ol R_{11} = -\partial_r \tau + \kappa\tau -|\chi|^2$,
we obtain
\begin{eqnarray*}
0 &=& - 2  (\partial_r + \frac{1}{2}\tau +  \kappa - \nu^0\partial_r \nu_0 )\tilde S
   + \tau \nu^0(\partial_r   + \frac{1}{2}\tau + 2 \kappa - 2\nu^0\partial_r \nu_0  )\ol A_{00}
 \;.
\end{eqnarray*}
We employ \eq{15III13.1c},
\begin{eqnarray*}
 \tilde S &=& -\nu^0(\partial_r   + \frac{1}{2}\tau  + 2 \kappa - 2\nu^0\partial_r \nu_0  )\ol A_{00}
 \;,
\end{eqnarray*}
which holds since all the hypotheses of Lemma~\ref{L19IV13.2} are fulfilled,
to end up with
\begin{eqnarray*}
  (\partial_r + \tau +  \kappa - \nu^0\partial_r \nu_0 )\tilde S &=& 0
 \;.
\end{eqnarray*}
Regularity at $O$ in coordinates as in~\cite[Section~4.5]{CCM2} gives $\tilde S = O(r^{-1})$, which implies that $\tilde S=0$ is the only possibility.
\qed

\subsection{Analysis of the KID equations in some special cases}
 \label{s9IV13.1}

\subsubsection{KID equations}

Theorem~\ref{main_result} shows that a vacuum spacetime emerging as solution of the characteristic initial value problem with data on a light-cone
possesses a Killing field if and only if the conformal class $\gamma_{AB}=[\ol g_{AB}]$ of $g_{AB}$, which together with $\kappa$ describes the free data on the light-cone, is such that,
 in the region where $\tau$ has no zeros, the \textit{KID  equations}
$\ol A_{ij}=0$ admit a non-trivial solution $\ol Y$. Written as equations for the vector field $\ol Y$, they read
(we use the formulae from~\cite[Appendix~A]{CCM2})
\begin{eqnarray}
  (\partial_r  - \kappa + \nu^0\partial_r \nu_0 ) \ol Y^0&=&0
\;,
\label{coneKID1}
\\
 \partial_r  \ol Y^A +(  \tilde\nabla^A\nu_0  -\partial_r \ol g^{1A}+ \kappa \nu^A
 + \nu_0\xi^A + \nu_0 \tilde\nabla^A) \ol Y^0 &=& 0
\;,
\label{coneKID2}
\\
 \tau \ol Y^1   + \tilde\nabla_A\ol Y^A -  \frac{1}{2}\nu_0(\zeta +\tau g^{11}+2\ol g^{1A} \tilde\nabla_A )\ol Y^0
 &=& 0
\;,
\label{coneKID3}
\\
(\tilde\nabla_{(A}\ol Y_{B)})\breve{} +\sigma_{AB}\ol Y^1  - \nu_0(\ol g^{11}  \sigma_{AB} + \breve{\ol \Gamma}{}^{1}_{AB} ) \ol Y^{0} &=& 0
\;,
\label{coneKID4}
\end{eqnarray}
where $\sigma_A{}^B$ denotes the trace-free part of $\chi_A{}^B$, $\xi^A:=\ol g^{AB}\xi_B$ and
%
\bea
\zeta &:= & 2\ol g^{AB}\ol\Gamma^1_{AB} + \tau \ol g^{11}
 \;, \label{11V13.2}
\\
 \xi_A &:= & -2\ol \Gamma^1_{1A}
 \;.
 \phantom{xx}
 \label{11V13.3}
\eea
The analysis of these equations is identical to that of their covariant counterpart, already discussed in Section~\ref{sec_free_data}.
The first three equations, arising from $\ol A_{1i}=0$ and $\ol g^{AB} \ol A_{AB}=0$ determine a class of candidate fields
(depending on the integration functions $c(x^A)$ and $f_A(x^B)$, with $\mathring\mcD_A c$ and $f_A$ being conformal Killing fields on $(S^{n-1},s)$. Note that it is crucial for the expansion $\tau$ to be non-vanishing in order for $\ol g^{AB} \ol A_{AB}=0$
to provide an algebraic equation for $\ol Y^1$.
Regardless of whether $\tau $ has zeros or not, we can determine $\ol Y^1$ by integrating radially~\eq{11V13.11}, compare Remark~\ref{R13V13.1}  below.

\subsubsection{Killing vector fields tangent  to spheres}

Let us consider the special case where the spacetime admits a Killing field $X$ with the property that $\ol X^0 = \ol X^1 =0$ on $C_O$.
The KID equations for the candidate field $\ol Y$  \eq{coneKID1}-\eq{coneKID4} then reduce to
\begin{eqnarray*}
 \partial_r  Y^A &=& 0
 \;,
\\
 \tilde\nabla_{(A} Y_{B)} &=& 0
 \;,
\end{eqnarray*}
which leads us to the following corollary of Theorem~\ref{main_result}:

\begin{corollary}
 \label{C10V13.1}
Consider initial data
$ \bar g_{AB}(r,x^C) dx^A dx^B$ for the vacuum Einstein equations (cf., e.g.,~\cite{ChPaetz}) on a light-cone $C_O$. In the resulting vacuum spacetime there exists a Killing field $X$ with  $\ol X^0 = \ol X^1 =0$ on $C_O$ defined on  a neighborhood of the vertex $O$
if and only if the family of Riemannian manifolds
$$
 (S^{n-1},g_{AB}(r,\cdot)\,dx^A dx^B)
$$
admits an $r$-independent Killing field $f^A=f^A(x^B)$.
\end{corollary}

\subsubsection{Killing  vector fields tangent  to the light-cone}
 \label{ss15VII13.1}

Let us now restrict attention to those Killing fields which are tangent  to the cone $C_O$, i.e.\ we assume
\begin{equation}
 \overline X{}^0 \,=\, 0
 \;.
  \label{10IV13.3}
\end{equation}
We start by noting that in the coordinates of \eq{1IX.1} we have
$$
 \underline X_\mu = \omega_{\mu\nu} y^\nu + O(|y|^2)
 \;,
$$
for an anti-symmetric matrix $\omega_{\mu\nu}$. Hence, quite generally,
\bean
  \displaystyle
   { {X}} ^0 & = & \frac{\partial x^0}{\partial y^\mu} \underline X^\mu
    = -  \underline X^0 + \frac {y^i}r  \underline X^i
    = \omega_{0i}y^i  + \frac {y^i}r (\omega_{ij}y^j-\omega_{0i}y^0)   + O(|y|^2)
     \nonumber
\\
     &= &   O(|y|^2)
  \;,
   \label{10IV13.1}
\\
  \displaystyle
   { {X}} ^1
    & = &
      \frac{\partial r}{\partial y^\mu} \underline X^\mu
      =  \omega_{i0}\frac {y^i}r y^0    + O(|y|^2)
    \;.
\eeal{10IV13.2}
Thus \eq{10IV13.3} does not impose any restrictions on $\omega_{\mu\nu}$, and we have
\bea
  \displaystyle
  \ol { {X}} ^1
    & = &  \omega_{i0} {y^i}     + O(r^2)
    \;.
\eeal{10IV13.4}
Next, under \eq{10IV13.3} the KID equations \eq{coneKID1}-\eq{coneKID4} for the candidate field $\ol Y$ become
\begin{eqnarray}
 \partial_r  \ol Y^A  &=& 0
\;,
\label{coneKID2_tang}
\\
 \tau \ol Y^1   + \tilde\nabla_A\ol Y^A  &=& 0
\;,
\label{coneKID3_tang}
\\
(\tilde\nabla_{(A}\ol Y_{B)})\breve{} +\sigma_{AB}\ol Y^1 &=& 0
\;,
\label{coneKID4_tang}
\end{eqnarray}
or, equivalently (note that $\partial_r \tilde\Gamma^B_{AB}= \partial_A\tau$)
\begin{eqnarray}
 \label{13V13.22}
 \ol Y^A &=& f^A(x^B)
 \;,
\\
 \label{13V13.23}
\partial_r (\tau \ol Y^1) + f^A\partial_A\tau &=& 0
 \;,
\\
 \tilde\nabla_{(A} f_{B)}   +\chi_{AB}\ol Y^1 &=& 0
\;,
 \label{13V13.21}
\end{eqnarray}
where we have set $f_A := \ol g_{AB}f^B$.  \Eqs{13V13.22}{13V13.21}
provide  thus a relatively simple form of the necessary-and-sufficient conditions for existence of Killing vectors tangent  to $C_O$.

If we choose a gauge where  $\tau =(n-1)/r$ (cf.\ e.g.\ \cite{ChPaetz}), the last three equations become
\begin{eqnarray}
 \ol Y^A &=& f^A(x^B)
 \;,
\\
 \ol Y^1 &=& -\frac{r}{n-1}\tilde\nabla_A f^A = -\frac{r}{n-1}\mathring \mcD_A f^A
 \;,
  \label{31VII13.11}
\\
 (\tilde\nabla_{(A} f_{B)})\breve{}   &=& \sigma_{AB}\frac{r}{n-1}\mathring \mcD_C f^C
\;.
\end{eqnarray}
Note that there are no non-trivial Killing vectors tangent to all generators of the cone, $\ol Y^A=0$, as \eq{31VII13.11} gives then $\ol Y^1=0$. This should be contrasted with a similar question for intersecting null hypersurfaces, see Section~\ref{ss14VII13.1}.

\section{Two intersecting null hypersurfaces}
 \label{s18III13.2}

\subsection{An intermediate result}

In analogy with the light-cone-case let us first prove an intermediate result:
\begin{Theorem}
 \label{hyp_intermediate}
Consider two smooth null hypersurfaces $N_1=\{x^1=0\}$ and $N_2=\{x^2=0\}$ in an $(n+1)$-dimensional  vacuum spacetime $(\mcM,g)$,
with transverse intersection along a smooth submanifold $S$.
Let $\ol Y$ be a  vector field defined on $N_1\cup N_2$. There exists  a smooth  vector field $X$ satisfying the Killing equations on $D^+(N_1\cup N_2)$ and coinciding with $\ol Y$ on $N_1\cup N_2$  if and only if \underline{on $N_1$} it holds that
\begin{eqnarray}
&
\coneD_2 \overline{Y}_2 =0
\label{hyp_cond1_int}
 \;,
&
\\
&
 \coneD_{(2} \ol Y_{A)}  =0
\label{hyp_cond2_int}
 \;,
&
\\
&
 \coneD_{(A} \ol Y_{B)} =0
\label{hyp_cond3_int}
 \;,
&
\\
 &
  R{}_{122}{}^{\mu} \ol Y_{\mu}-\coneD_{2}\coneD_2\ol Y_1 = 0
\label{hyp_cond4_int}
  \;,
&
\\
 &
  R{}_{12A}{}^{\mu} \ol Y_{\mu}-\coneD_{A}\coneD_2\ol Y_1 = 0
\label{hyp_cond5_int}
  \;,
&
\\
 &
  g^{AB} (R{}_{1AB}{}^{\mu} \ol Y_{\mu}-\coneD_{A}\coneD_B\ol Y_1 )= 0
\label{hyp_cond6_int}
  \;,
&
\end{eqnarray}
where $\coneD$ is the analogue on  $N_1$ of the derivative operator \eq{26III13.11}-\eq{26III13.21}; similarly on $N_2$; while on $S$ one needs further to assume that
\begin{eqnarray}
 &
 (\coneD_1 Y_{2} + \coneD_{2}Y_1)|_S = 0
 \;.
 \label{hyp_cond_S_int}
\end{eqnarray}
\end{Theorem}

{\noindent\sc Proof:}
The proof is essentially identical to the proof of Theorem~\ref{T18I13.1}. The candidate field is constructed as a solution of the wave equation
\eq{20XI12.2}; the delicate question of regularity of $\ol Y$ needed at the vertex in the cone case does not arise.  Existence of the solution in $J^+(N_1\cup N_2)$ follows from~\cite{RendallCIVP}.

The main difference is that one cannot invoke regularity at the vertex to deduce
the vanishing of, say on $N_2$, $A_{2\mu}|_{N_2}$ from the equations which correspond to \eq{15III13.1}, \eq{15III13.1b} and \eq{15III13.1c}.
Instead, one needs further to require \eq{hyp_cond_S_int} as well as
\begin{eqnarray*}
 (\nabla_2 Y_{A} + \nabla_{A}Y_2)|_S = 0 \quad \text{and} \quad
 \nabla_{2}Y_2|_S = 0
 \;.
\end{eqnarray*}
However, the last two conditions follow from \eq{hyp_cond1_int} and \eq{hyp_cond2_int} on $N_1$.
\qed

\subsection{Proof of Theorem~\ref{T18I13.2}}
\label{S_hyp_proof_main_result}

We prove now our main result for transversally intersecting null hypersurfaces:

\vspace{0.5em}
{\noindent \sc Proof of Theorem~\ref{T18I13.2}:}
We want to show that \eq{hyp_cond1}-\eq{hyp_cond2b} imply that all the remaining
assumptions of Theorem~\ref{hyp_intermediate}, namely
\eq{hyp_cond1_int}-\eq{hyp_cond_S_int}, are  satisfied.
The conditions \eq{hyp_cond1_int}, \eq{hyp_cond2_int}, \eq{hyp_cond4_int} and \eq{hyp_cond_S_int} follow trivially.

Lemma~\ref{lem_relation_trA_S110}, adapted to the intersecting null hypersurfaces-setting, tells us that $g^{AB}\nabla_{(A}Y_{B)}$ vanishes on
$N_1\cup N_2$ due to \eq{hyp_cond2a} and \eq{hyp_cond2aa}.
Hence \eq{hyp_cond3_int} is fulfilled.

The analogue of Lemma~\ref{SA10_vanishes} for two intersecting null hypersurfaces requires, in addition to \eq{hyp_cond2b},  the vanishing of
\begin{eqnarray}
\coneD_{A}\coneD_{(1} \ol Y_{2)} |_S = 0
 \;.
\label{additional condition}
\end{eqnarray}
Both  \eq{hyp_cond2b} and \eq{additional condition} together imply vanishing initial data for the analogue of the ODE \eq{ODE_SA10} in the current setting.
Equation \eq{additional condition} follows from \eq{hyp_cond2}, the corresponding equation on $N_2$, and \eq{hyp_cond_S}.
Thus \eq{hyp_cond5_int} is fulfilled.

A straightforward adaptation of Lemma~\ref{gAB_SAB0_vanishes} to the current setting shows that, say on $N_2$, $g^{AB}S_{AB2}-\frac{1}{2}\tau g^{12}A_{22}$ vanishes, supposing that it vanishes on $S$.
Using Lemma~\ref{properties_S} (i) we find
\begin{eqnarray*}
 (g^{AB}S_{AB2}-\frac{1}{2}\tau g^{12}A_{22})|_S &=& g^{AB}(\nabla_{(A}A_{B)2} -  \frac{1}{2} \nabla_2 A_{AB})-\frac{1}{2}\tau g^{12}A_{22}
\\
&=& 0
\;,
\end{eqnarray*}
because of \eq{hyp_cond1}-\eq{hyp_cond2aa}.
Hence also \eq{hyp_cond6_int} is fulfilled.
\qed

\begin{remark}
 \label{R13V13.1}
{\rm
While $\tau$ has no zeros near the tip of a light-cone, for two transversally intersecting null hypersurfaces the expansion $\tau$ may vanish even near the intersection.
In that case the trace of \eq{hyp_cond3_int} on, say, $N_1$ will fail to provide an algebraic equation for  $\ol X^2$. Also, Corollary~\ref{S110_vanishes} cannot be applied to deduce the vanishing of $\ol S_{221}$, equivalently,  the validity
of \eq{hyp_cond4_int}, in the regions where $\tau$ vanishes. Instead one can use the second-order ODE  \eq{hyp_cond4_int} to find a candidate for $\ol X^2$,
and then Lemma~\ref{lem_relation_trA_S110} guarantees that the trace of the left-hand side of \eq{hyp_cond3_int} vanishes when  $g^{AB}A_{AB}|_S=0=\partial_2(g^{AB}A_{AB})|_S$.
}
\end{remark}

\subsection{Bifurcate horizons}
 \label{ss14VII13.1}

A key notion to the understanding of the geometry of stationary black holes is that of a \emph{bifurcation surface}. This is a smooth submanifold $S$ of co-dimension two on which a Killing vector $X$ vanishes, with $S$ forming a transverse intersection of two smooth null hypersurfaces so that $X$ is tangent to the generators of each. In our context this would correspond to a KID which vanishes on $S$, and is tangent to the null generators of the two characteristic hypersurfaces emanating normally from $S$. In coordinates adapted to one of the null hypersurfaces, so that the hypersurface is given by the equation $x^1=0$, we have $\overline X = \ol X^ 2 \partial_2$, and $\ol X^ \flat = \ol g_{12} \ol X^ 2 dx^ 1$.
 Then \eq{10IV13.3} holds, and therefore also \eq{coneKID2_tang}-\eq{coneKID4_tang}
(which correspond to \eq{hyp_cond2_int} and \eq{hyp_cond3_int}). Equations~\eq{coneKID3_tang}-\eq{coneKID4_tang} show that this is only possible if $\tau=\sigma_{AB}=0$, which implies that translations along the generators of the light-cone are isometries of the $(n-1)$-dimensional metric $\ol g_{AB} dx^ A dx^ B$. Equivalently, $N_1$ and $N_2$ have vanishing null second fundamental forms, which provides a necessary condition for a bifurcate horizon.

Assuming vacuum (as everywhere else in this work),
this condition turns out to be sufficient. Let $\zeta_A$ be the torsion one-form of $S$ (see, e.g.,~\cite{Ch-KL}, or \eq{20VII13.14} below). We can use Theorem~\ref{T18I13.1} to prove (compare~\cite[Proposition~B.1]{FRW} in dimension $3+1$ and \cite[end of Section~2]{HIW} in higher dimensions):

\begin{Theorem}
 \label{T15VII13.1}
 Within the setup of Theorem~\ref{T18I13.1}, suppose that the null second fundamental forms of the hypersurfaces $N_a$, $a=1,2$, vanish.
 Then:
 \begin{enumerate}
   \item
There exists a Killing vector field $X$ defined on $D^+(N_1\cup N_2)$  which vanishes on $S$ and is null on $N_1
 \cup N_2$.
 \item
 Furthermore, any Killing vector $\hat Y = \hat Y^A\partial_A$ of the metric induced by $g$ on $S$
 extends to a Killing vector $X$ of $g$ on $D^+(N_1\cup N_2)$ if and only if the $\hat Y$-Lie derivative
of the torsion one-form of $S$ is exact.
   \end{enumerate}
\end{Theorem}

\begin{remarks}{\rm

  1.
Killing vectors as above would exist to the past of $S$ if the past-directed null hypersurfaces emanating from $S$ also had vanishing second null fundamental forms. However,
this does not need to be the case, a vacuum example is provided by suitable Robinson-Trautman spacetimes.

2. Concerning point 2.\ of the Theorem, when the $\hat Y$-Lie derivative of $ \zeta$ is merely closed the argument of the proof below provides one-parameter families of  Killing vectors defined on  domains of dependence $D^+(\mcO)$ of  simply connected subsets $\mcO$ of $S$.
It would be of interest to find out whether or not the resulting locally defined Killing vectors can be patched together to a global one  when $S$ is not simply connected.
 }
\end{remarks}

 \medskip

\proof
1. In coordinates adapted to the null hypersurfaces the condition that a Killing vector $  X$ is tangent to the generators is equivalent to
\begin{eqnarray}
&&   X^{\mu}|_S=0\;,\quad   X^1|_{N_1}=  X^2|_{N_2}=0\;,  \quad   X^A|_{N_1\cup N_2}=0
 \;,
 \label{KVF_initial}
 \\
\hspace{-2em}  \Longleftrightarrow &&
   X_{\mu}|_S=0\;,\quad   X_2|_{N_1}= X_1|_{N_2}=0\;,  \quad    X_A|_{N_1\cup N_2}=0
  \;.
\end{eqnarray}
For simplicity we assume that the generators of the
two null hypersurfaces are affinely parameterized, i.e.\ $\kappa_{N_1}=\kappa_{N_2}=0$. By hypothesis we have
\begin{eqnarray}
&
 \tau_{N_1} = \tau_{N_2} =\sigma^{N_1}_{AB}= \sigma_{AB}^{N_2}  =0
 \label{shear_equation}
 \;.
&
\end{eqnarray}
The KID equations \eq{hyp_cond1}-\eq{hyp_cond2b} for the candidate field $\ol Y$ reduce to
\begin{eqnarray}
 &
 \partial_2\partial_2\ol Y_1  -2\Gamma^{1}_{12}\partial_2\ol Y_{1}
 + ( (\Gamma^{1}_{12})^2  -\partial_2\Gamma^{1}_{12})\ol Y_{1}|_{N_1}   =0
 \;,
  \label{ODE_N1}
 &
 \\
 &
 \partial_1\partial_1\ol Y_2  -2\Gamma^{2}_{12}\partial_1\ol Y_{2}
 + ( (\Gamma^{2}_{12})^2  -\partial_1\Gamma^{2}_{12})\ol Y_{2}|_{N_2}   =0
   \label{ODE_N2}
 \;,
 &
\\
 &
 (\partial_1\ol  Y_{2} + \partial_{2}\ol Y_1)|_S = 0
 \;,
 &
\\%
 &
 \partial_A (\partial_1 \ol Y_2 - \partial_2 \ol Y_1 )|_S = 0
   \;.
 &  \label{ODE_N4}
\end{eqnarray}
Since $\ol Y_{\mu}|_S=0$ we need non-trivial initial data $\partial_2\ol Y_1|_{S}$ and $\partial_1\ol Y_2|_{S}$
for the ODEs \eq{ODE_N1} and \eq{ODE_N2} for $\partial_2\ol Y_1|_{N_1}$ and $\partial_1\ol Y_2|_{N_2}$,
respectively.

Using the formulae in \cite[Appendix~A]{CCM2}, \eq{ODE_N1}-\eq{ODE_N4} can be rewritten as
\begin{eqnarray}
 &
 \partial_2\partial_2\ol Y^2|_{N_1}  =0
 \;,
  \label{ODE_N1b}
 &
 \\
 &
 \partial_1\partial_1\ol Y^1|_{N_2}  =0
   \label{ODE_N2b}
 \;,
 &
\\
 &
 (\partial_1 \ol Y^1 + \partial_{2}\ol Y^2)|_S = 0
 \;,
\label{sum_on_S}
 &
\\
 &
 \partial_A(\partial_1 \ol Y^1- \partial_2 \ol Y^2)|_S = 0
\label{KID_bif}
   \;.
 &
\end{eqnarray}
Hence there remains the freedom to prescribe  a constant $c\ne 0$ for $\frac{1}{2}(\partial_1 \ol Y^1- \partial_2 \ol Y^2)|_S $.
(The constant $c$ reflects the freedom of scaling the Killing vector field by a constant, and is related to the surface
gravity of the horizon; we will return to this shortly.)
By  \eq{sum_on_S} one needs to choose $\partial_1\ol  Y^1|_S=c $ and $\partial_2\ol  Y^2|_S=-c $. Together with $\ol  Y^1|_S=\ol  Y^2|_S=0$
the functions $\ol  Y^2|_{N_1}$ and $\ol  Y^1|_{N_2}$ are then determined by
\eq{ODE_N1b} and \eq{ODE_N2b}, and existence of the desired Killing vector follows from Theorem~\ref{T18I13.1}.

2. By assumption we have, using obvious notation,
\begin{equation}
 \chi^{N_1}_{AB} = \chi^{N_2}_{AB} =0 \quad \Longleftrightarrow \quad
\partial_2 g_{AB}|_{N_1} =  \partial_1 g_{AB}|_{N_2} =0
 \;.
\label{chi_vanishing}
\end{equation}
Now, the flow of a space-time Killing vector field which is tangent to $S$ preserves $S$. This implies that the bundle
of null vectors normal to $S$ is invariant under the flow. Equivalently, the image by the flow of a null geodesic normal to $S$ will be a one-parameter family of null geodesics normal to $S$. This is possible only if the Killing vector field is tangent to both $N_1$ and $N_2$.
It thus suffices to consider candidate Killing vector fields $\ol Y$
 which satisfy, in our adapted coordinates,
\begin{equation}
 \ol Y^1|_{N_1}=\ol Y^2|_{N_2}=0 \quad \Longleftrightarrow \quad \ol Y_2|_{N_1}=\ol Y_1|_{N_2}=0
\;.
 \label{tangent_cond}
\end{equation}

To continue, we need a simple form of the KID equations \eq{hyp_cond1}-\eq{hyp_cond2b}, assuming \eq{chi_vanishing} and \eq{tangent_cond}, and supposing again that the generators of the
two null hypersurfaces are affinely parameterized, i.e.\ $\kappa_{N_1}=\kappa_{N_2}=0$.
Using the notation $\hat Y \equiv Y^A|_S\partial_A$ we find:
\begin{eqnarray}
&
 \partial_{2}\ol  Y^{A}|_{N_1} =0
 \;,
 \quad
 \partial_{1} \ol Y^{A}|_{N_2} =0
  \;,
\label{tangent_KID_1}
&
\\
&
( \tilde\nabla_{(A}  \ol Y_{B)})\breve{}\,|_{N_1\cup N_2} =0
 \;,
\label{tangent_KID_2}
&
\\
 &
 \partial_2\partial_2 \ol Y^{2} |_{N_1}= 0\;, \quad   \partial_1\partial_1\ol  Y^{1} |_{N_2}= 0
  \;,
\label{tangent_KID_3}
&
\\
 &
 (\partial_1 \ol Y^1 + \partial_{2}\ol Y^2 +g^{12}  \mcL_{\hat Y} g_{12} )|_S = 0
 \;,
\label{tangent_KID_4}
 &
\\
&
 \tilde\nabla_A \ol Y^A    |_S =0
\;,
\label{tangent_KID_5}
&
  \\
 &
 [\partial_A (\partial_1\ol Y^1 - \partial_2\ol Y^2)  + 2\mcL_{\hat Y}\zeta_A]|_S = 0
\label{tangent_KID_6}
 &
\end{eqnarray}
(The fields $g_{12}|_S$ and
\begin{equation}
 \zeta_A = \frac{1}{2}(\Gamma^1_{1A}-\Gamma^2_{2A})|_S
\label{dfn_zeta_A}
\end{equation}
are part of the free initial data on $S$ \cite{RendallCIVP}; compare~\cite{ChPaetz}.)
The first-order equations above are straightforward; some details of the derivation of the remaining equations above will be given shortly. Before passing to that, we observe that  \eq{tangent_KID_1}-\eq{tangent_KID_2} and \eq{tangent_KID_5}, together with \eq{chi_vanishing},
are equivalent to the requirement that
$\hat Y|_S$ is a Killing vector field of $(S, g_{AB}|_S)$, and that $\ol Y^A = \hat Y^A$ on $N_1\cup N_2$, i.e.\ that $\ol Y^A$
is independent of the coordinates $x^1$ and $x^2$.
Supposing further that $\mcL_{ \hat Y} \zeta_A$ is exact, the remaining equations \eq{tangent_KID_3},  \eq{tangent_KID_4}  and  \eq{tangent_KID_6} can be used to determine $\ol Y^1$ and $\ol Y^2$ on $N_1\cup N_2$. (As such, on each  connected component of $S$ the difference $(\partial_1\ol Y^1 - \partial_2\ol Y^2)|_S$ is determined up to an additive constant by \eq{tangent_KID_6}, which reflects the freedom of adding a Killing vector field which vanishes on $S$ and is tangent to the null geodesics generating both initial surfaces.) The existence of a Killing vector field $X$ on the space-time which coincides with $\hat Y$ on $S$ follows now from Theorem~\ref{T18I13.2}.

In Appendix~\ref{gauge-dependence} it is shown that the KID equations \eq{tangent_KID_1}-\eq{tangent_KID_6} are invariant under affine reparameterizations of generators. We also show there that the freedom to choose an affine parameter on $N_1$ and $N_2$ can be employed to prescribe
$g_{12}$ on $S$, and also to add arbitrary gradients to $\zeta$.
In particular exactness of $\zeta$,
and thereby solvability of the KID equations, is independent of the gauge, as one should expect.

Let us pass to some details of the derivation of the second-order equations above.
The calculation uses extensively
$$
 0=\Gamma^\mu_{22}|_{N_1} =\Gamma^A_{B2}|_{N_1}
 \;,
$$
similarly on $N_2$.
First, let us compute $R_{122}{}^A|_{N_1}$ and $R_{211}{}^A|_{N_2}$, as needed to evaluate \eq{hyp_cond4}. Making use of  \cite[Appendix~A]{CCM2}, we find
\begin{eqnarray*}
 \partial_1\Gamma^A_{22}|_{N_1} &=& g^{2A}\partial_2\partial_2g_{12} + g^{AB}\partial_2(\partial_1g_{2B} - \partial_Bg_{12})
\\
&=& g^{2A}\partial_2\partial_2g_{12} + \partial_2(g_{12}g^{AB}\xi^{N_1}_B)
 + g^{AB}\partial_2 \partial_2 g_{1B}
 \;,
\\
\Gamma^A_{12}|_{N_1} &=& \frac{1}{2}g_{12}g^{AB}\xi^{N_1}_B + g^{AB}\partial_2 g_{1B} + g^{2A}\partial_2 g_{12}
 \;,
\end{eqnarray*}
and thus
\begin{eqnarray*}
 R_{122}{}^A|_{N_1}
 &=& (\partial_2 +g^{12}\partial_2g_{12} )\Gamma^A_{12}  - \partial_1\Gamma^A_{22}
\,=\,  \frac{1}{2}g_{12}g^{AB} \partial_2\xi^{N_1}_B
\;,
\end{eqnarray*}
where, using the notation in~\cite{ChPaetz}, $\xi^{N_1}_A=-2\Gamma^2_{2A}|_{N_1}$ and $\xi^{N_2}_A=-2\Gamma^1_{1A}|_{N_2}$.
Now for $\chi^{N_1}_{AB}=\chi^{N_2}_{AB}=0$ the vacuum characteristic constraint equations \cite{CCM2} imply $\partial_2\xi^{N_1}_A=0$ and $\partial_1\xi^{N_2}_A=0$,
hence
\begin{equation*}
 R_{122}{}^A|_{N_1}=R_{211}{}^A|_{N_2}=0
 \;.
\end{equation*}
Further simple calculations lead to \eq{tangent_KID_3}.

Next, again on $S$, we consider the KID equation
\bel{28VII13.10}
0= \nabla_A \nabla_1 \ol Y^1 - R_{\mu A 1}{}^1 \ol Y^\mu = \nabla_A \nabla_1 \ol Y^1 + R^1{}_{1  BA} \ol Y^B
 \;.
\ee
\underline{On $S$} it holds that
\beaa
 \nabla_1 \ol Y^\mu \partial_\mu &= & \partial_1 \ol Y^1 \partial_1 + \Gamma^\mu_{1 A} \ol Y^A \partial_\mu
 \;,
\\
 \nabla_\mu \ol Y^1 dx^\mu &= & \partial_1 \ol Y^1 dx^1+ \Gamma^1_{\mu A}\ol  Y^A dx^ \mu
 \;,
\\
 \nabla_A \nabla_1\ol  Y ^1
  & = &
   \partial_A (\partial_1 \ol Y^1 + \Gamma^1_{1B} \ol Y^B ) +\Gamma^1_{A\mu} \nabla_1 \ol Y^\mu
   - \Gamma^ \mu_{A1} \nabla_\mu \ol Y^1
\\
%
  & = &
   \partial_A \partial_1 \ol Y^1 + \Gamma^1_{1B} \partial_A \ol Y^B
    + \underbrace{\big(\partial_A \Gamma^1_{1B}   + \Gamma^1_{A\mu}\Gamma^\mu_{1 B}
   - \Gamma^ \mu_{A1} \Gamma^1_{\mu B}\big)}_{=\partial_B \Gamma^1_{1A} - R^1{}_{1BA} } \ol Y^B
   \;.
\eeaa
%
%
Inserting into \eq{28VII13.10} gives the following equation on $S$, in coordinates adapted to $N_2$:
\bean
 0
&=&
  \nabla_A \nabla_1 \ol Y^1 - R_{\mu A 1}{}^1 \, \ol Y^\mu
\\
  & = &
   \partial_A \partial_1 \ol Y^1 + \Gamma^1_{1B} \partial_A\ol  Y^B
    +\partial_B \Gamma^1_{1A}
    \ol Y^B
  \;.
\eeal{28VII13.11}
The analogous formula in coordinates adapted to $N_1$ reads
\bean
0 &=&
  \nabla_A \nabla_2 \ol Y^2 - R_{\mu A 2}{}^2\, \ol Y^\mu
\\
  & = &
   \partial_A \partial_2 \ol Y^2 + \Gamma^2_{2B} \partial_A\ol  Y^B
    +\partial_B \Gamma^2_{2A}
  \ol  Y^B
  \;.
\eeal{28VII13.12}
%
Subtracting we obtain \eq{tangent_KID_6}.

From the discussion so far it should be clear that the conditions are necessary. This concludes the proof.
\qed

As an example, suppose that $\hat Y$ is a Killing vector on $S$ and that the torsion one-form is invariant under the flow of $\hat Y$.
It follows from the equations above that we can reparameterise the initial data surfaces so that $g_{12}=1$ on $S$, with the torsion remaining invariant in the new gauge. Then
$\ol Y^1=\ol Y^2=0$ and $\ol Y^A = \hat Y^A$
provides a solution of the KID equations on $N_1\cup N_2$.

It is of interest to relate the constant $c$, arising in the paragraph following \eq{KID_bif}, to the surface gravity (which we denote by  $\kappa_\mcH$ here);
this will also prove in which sense the seemingly coordinate-dependent derivatives $\partial_1 X^1|_S=-\partial_2X^2|_S$ are in fact geometric invariants. In the process we recover the well-known fact, that surface gravity is constant on bifurcate horizons. We have
\begin{equation}
 \kappa_\mcH^2 = - \frac{1}{2}(\nabla^{\mu}X^{\nu})(\nabla_{\mu}X_{\nu})|_{N_1\cup N_2}
 = -(\nabla_1X^1)(\nabla_2X^2)|_{N_1\cup N_2}
\;.
\end{equation}
On,  say, $N_1$ we have due to \eq{KVF_initial} and \eq{ODE_N1b}-\eq{KID_bif}
\begin{equation}
 \nabla_2X^2|_{N_1} = \partial_2X^2 =- c
 \;,
\end{equation}
while $\nabla_1X^1|_{N_1}$ can be computed from \eq{20XI12.2},
\begin{eqnarray*}
 \Box X^1|_{N_1} = -R^1{}_{\alpha}X^{\alpha} = -g^{12}R_{22}X^2 =0 && \Longleftrightarrow \quad
 \partial_2\nabla_1X^1|_{N_1} =0
\\
 && \Longleftrightarrow \quad
 \nabla_1X^1|_{N_1} =c
 \;,
\end{eqnarray*}
where we used \eq{KVF_initial}, the vanishing of $
\chi_{AB}$, and $\nabla_1X^1|_S=c$. Hence
\begin{equation}
 \kappa_\mcH = |c|
\;.
\end{equation}

\appendix

\section{Fuchsian ODEs}
 \label{A5VIII13.1}

Since it appears difficult to find an adequate reference, we describe here the main property of first-order Fuchsian ODEs used in our work.

Consider a first order system of equations for a set of fields $\phi=(\phi^I)$, $I=1,\ldots,N$, of the form
\bel{5VIII13.1}
 r \partial_r \phi = A(r) \phi + F(r,\phi)
 \;,
\ee
for some smooth map $F=(F^I)$ with $F(0,0)=0$, $\partial_\phi F(r,0)=0$, where $A(r)$ is a smooth map with values in $N\times N$ matrices.
For our purposes it is sufficient to consider the case where
$$
A(0) =\lambda \Id
\;,
$$
where $\Id$ is the $N\times N$ identity matrix.  It holds that \emph{the only solution of \eq{5VIII13.1} such that
$\lim_{r\to 0} r^{-\lambda}\phi(r)=0$ is $\phi(r)=0$ for all $r$.}

\section{Gauge-dependence of the torsion one-form}
\label{gauge-dependence}

In this appendix we consider the question of gauge-independence in point 2 of Theorem~\ref{T15VII13.1}.
Indeed, even within the gauge conditions imposed so far, that $x^1$ and $x^2$ are affine parameters on the relevant characteristic surfaces, there is some gauge freedom left concerning the gravitational initial data.
The point is that we can rescale the affine parameters $x^2$ on $N_1$ and $x^1$ on $N_2$,
\bel{20VII13.11}
 x^2  \mapsto  \check x^2 = e^{-f^+(x^B)} x^2
  \;,
  \quad
 x^1  \mapsto  \check x^1 = e^{-f^-(x^B)} x^1
 \;,
\ee
with some functions $f^\pm$ defined on $S$.
Under \eq{20VII13.11}, the metric on $N_1$ becomes
\bean
    g|_{N_1}
     & = &
      \ol g_{11} (dx^1)^2 + 2 (\ol g_{12} dx^2 + \ol g_{1A} dx^A)dx^1 + \ol g_{AB}dx^ A dx^ B
\\
     & = &
      e^{2f^-}\ol g_{11}(d\check x^1)^2 + 2 \big( e^{f^+}\ol g_{12}  d\check x^2 +  (\ol g_{1A} + \check x^2 e^{f^+}\ol g_{12}\partial_Af^+)dx^A\big)e^{f^-}d\check x^1
      \nonumber
\\
     & = &
      e^{2f^-}\ol g_{11}(d\check x^1)^2 + 2 \big(\underbrace{e^{(f^++f^-)}\ol g_{12}}_{\ol{\check g}_{12}} d\check x^2 +
       \underbrace{e^{f^-}(\ol g_{1A} +   x^2  \ol {  g}_{12}\partial_Af^+)}_{=:\ol{\check g}_{1A}}dx^A\big)d\check x^1
\nonumber
\\
 &&
      + \ol g_{AB}dx^ A dx^ B \;,
\eeal{20VII13.12}
with a similar formula on $N_2$. This leads to
\bel{20VII13.13}
 \check g_{12}|_S=e^{(f^++f^-)} g_{12}|_S
 \;,
\ee
as well as, using \cite[Equation~(2.12)]{ChPaetz},
\bean
 \check \zeta_A
  &= &
    \frac 12 \check g^{12}
  \big(
   \frac{\partial \check g_{1A}}{\partial \check x^2}
  -
   \frac{\partial \check g_{2A}}{\partial \check x^1}
    \big)\big|_S
\\
 \nonumber
  &
     =
   &
    \frac 12 e^{-(f^++f^-)} g^{12}
  \big(
    e^{f^+} \frac{\partial \big(e^{f^-}(\ol g_{1A} +   x^2 \ol g_{12}\partial_Af^+) \big) }{\partial   x^2}
  -
   \frac{\partial \check g_{2A}}{\partial \check x^1}
    \big)\big|_S
\\
 &     =
  &
    \zeta_A + \frac 12
  \big(\partial_A f^+-\partial_Af^-
    \big)\big|_S
     \;.
\eeal{20VII13.14}
Letting $\check x^A = x^ A$, in the new coordinates the Killing vector becomes
\bean
 Y &= & Y^\mu \partial_{x^\mu} = Y^\mu \frac{\partial \check x^ \nu}{\partial x^\mu} \partial_{\check x^\nu}
 \\
 & = &
  Y^\mu \frac{\partial(e^{-f^-} x^1)}{\partial x^\mu} \partial_{\check x^1}
  +
   Y^\mu \frac{\partial (e^{-f^+} x^2)}{\partial x^\mu} \partial_{\check x^2}
   + Y^\mu \frac{\partial \check x^ A}{\partial x^\mu} \partial_{\check x^A}
   \nonumber
 \\
 & = &
  \big(
 e^{-f^-} Y^1 -
  \check  x^1 Y^A \partial_A f^-\big)  \partial_{\check x^1}
   \nonumber
 \\
 & &
  +
  \big(
 e^{-f^+} Y^2 -
  \check  x^2 Y^A \partial_A f^+\big)  \partial_{\check x^2}
   + Y^A   \partial_{\check x^A}
 \;.
\eeal{28VII13.3}
Invariance of \eq{tangent_KID_1}-\eq{tangent_KID_3} and \eq{tangent_KID_5} is clear.
One can further check invariance of \eq{tangent_KID_4} (recall that $\hat Y \equiv Y^A\partial_A|_S$):
\bean
\lefteqn{
(\partial_{\check x^1} \check Y^1 + \partial_{\check x^ 2}\check Y^2 +\check  g^{12}  \mcL_{\hat Y} \check g_{12} )|_S
 }
 &&
\\
 && =
\big(\partial_{\check x^1}  \big(
 e^{-f^-} Y^1 -
  \check  x^1 Y^A \partial_A f^-\big) + \partial_{\check x^ 2}\check Y^2+e^{-(f^++f^-)}g^{12}   \mcL_{\hat Y}( e^{f^++f^-}g_{12}) \big)|_S
 \nonumber
\\
 &&
  = (\partial_1 Y^1 + \partial_{2}Y^2 -\hat Y^A\partial_A(f^+ +f^-) +\hat Y^A\partial_A(f^++f^-)+g^{12}  \mcL_{ \hat Y} g_{12} )|_S
  \nonumber
\\
 &&
  = (\partial_1 Y^1 + \partial_{2}Y^2 +g^{12}    \mcL_{ \hat Y} g_{12} )|_S
 \;.
\label{tangent_KID_4check}
\eea

As such, \underline{on $S$} the first two-terms in \eq{tangent_KID_6} transform as
\bean
 \lefteqn{
 \partial_{\check x^A}(\partial_{\check x^1} \check Y^ 1 - \partial_{\check x^2} \check Y^2)
 =
 \frac{\partial x^\mu}{\partial \check x^A}
 \partial_\mu(\partial_{\check x^1} \check Y^ 1 - \partial_{\check x^2} \check Y^2)
 }
 &&
\\
 && \phantom{xxxx}=
  \partial _A
 \big(
  \partial_1 Y^1 - \partial_{2}Y^2 + \hat Y^B\partial_B(f^--f^+ )
   \big)
   \;.
\eeal{28VII13.4}
\Eq{20VII13.14} can be rewritten as
\bea
 \check \zeta=
    \zeta + \frac 12d
  ( f^+- f^-
    )
     \;.
\eeal{20VII13.14+}
Thus
\bean
 2\mcL_{\hat Y} \check \zeta = \mcL_{\hat Y} \big( 2\zeta +   d  ( f^+- f^-
    ) \big)
     =2 \mcL_{\hat Y}  \zeta +   d \big(\mcL_{\hat Y} ( f^+- f^-
    ) \big)
    \;,
\eeal{28VI13.7}
which shows that the one-form
\bel{28VI13.8}
 d (\partial_1 Y^1 - \partial_2Y^2) + 2 \mcL_{\hat Y} \zeta
\ee
is invariant under changes of the affine parameters, as desired.

We end this paper by deriving the behaviour of $\zeta_A\equiv \frac{1}{2}(\Gamma^1_{1A}-\Gamma^2_{2A})|_S$ under arbitrary coordinate transformations which preserve the adapted null coordinates conditions,
\begin{equation}
 \check x^1 = e^{-f^+(x^{\mu})}x^1\;, \quad   \check x^2 = e^{-f^-(x^{\mu})}x^2\;, \quad \check x^A = x^A
 \;.
\label{gen_coord_trafo}
\end{equation}
We set
\begin{equation*}
 f_0^{\pm}(x^A) := f^{\pm}(x^1=0,x^2=0,x^A)
 \;.
\end{equation*}
Then
\begin{eqnarray*}
 (\check\Gamma^1_{1A}-\check\Gamma^2_{2A})|_S
 &=& \Gamma^{\sigma}_{\mu\nu} \Big(\frac{\partial \check x^1}{\partial x^{\sigma} }\frac{\partial x^{\mu}}{\partial \check x^{1}}
- \frac{\partial \check x^2}{\partial x^{\sigma} }\frac{\partial x^{\mu}}{\partial \check x^{2}}\Big)
\frac{\partial x^{\nu}}{\partial \check x^{A}}
+  \frac{\partial \check x^1}{\partial x^{\sigma} } \frac{\partial^2 x^{\sigma}}{\partial \check x^{1}\partial \check x^{A}}
-  \frac{\partial \check x^2}{\partial x^{\sigma} } \frac{\partial^2 x^{\sigma}}{\partial \check x^{2}\partial \check x^{A}}
\\
&=& \Gamma^{1}_{1 B} \frac{\partial \check x^1}{\partial x^{1} }\frac{\partial x^{1}}{\partial \check x^{1}}
\frac{\partial x^{B}}{\partial \check x^{A}}
  - \Gamma^{2}_{2 B} \frac{\partial \check x^2}{\partial x^{2} }\frac{\partial x^{2}}{\partial \check x^{2}}
\frac{\partial x^{B}}{\partial \check x^{A}}
\\
&&
+  \frac{\partial \check x^1}{\partial x^{1} } \frac{\partial^2 x^{1}}{\partial \check x^{1}\partial \check x^{A}}
-  \frac{\partial \check x^2}{\partial x^{2} } \frac{\partial^2 x^{2}}{\partial \check x^{2}\partial \check x^{A}}
\\
&=& \Gamma^{1}_{1 A}e^{-f_0^+}\frac{\partial x^{1}}{\partial \check x^{1}}
  - \Gamma^{2}_{2 A} e^{-f_0^-}\frac{\partial x^{2}}{\partial \check x^{2}}
+ e^{-f_0^+} \partial_A\frac{\partial x^{1}}{\partial \check x^{1}}
-  e^{-f_0^-} \partial_A\frac{\partial x^{2}}{\partial \check x^{2}}
\\
&=& \Gamma^{1}_{1 A}
  - \Gamma^{2}_{2 A}
+\partial_A(f_0^+ -  f_0^-)
 \;,
\end{eqnarray*}
i.e.\  \eq{20VII13.14+} holds under coordinate transformations of the form \eq{gen_coord_trafo}.

If we assume $S$ to be compact, there is a natural way to fix the gauge: According to the Hodge Decomposition Theorem $\zeta$ can be uniquely
written as the sum of an exact one-form, a dual exact one-form and a harmonic one-form. The considerations above show that the first term has a pure gauge
character and can be transformed away, while the remaining part has an intrinsic meaning.
In particular, if $\zeta$ is exact, the remaining gauge freedom can be employed to transform it to zero.

\bibliographystyle{amsplain}
\bibliography{ConeKIDSNew-minimal}

\def\polhk#1{\setbox0=\hbox{#1}{\ooalign{\hidewidth
  \lower1.5ex\hbox{`}\hidewidth\crcr\unhbox0}}} \def\cprime{$'$}
  \def\cprime{$'$}
\providecommand{\bysame}{\leavevmode\hbox to3em{\hrulefill}\thinspace}
\providecommand{\MR}{\relax\ifhmode\unskip\space\fi MR }
\providecommand{\MRhref}[2]{%
  \href{http://www.ams.org/mathscinet-getitem?mr=#1}{#2}
}
\providecommand{\href}[2]{#2}
\begin{thebibliography}{10}

\bibitem{ChBeigKIDs}
R.~Beig and P.T. Chru\'{s}ciel, \emph{Killing {I}nitial {D}ata}, Class.\
  Quantum. Grav. \textbf{14} (1997), A83--A92, A special issue in honour of
  Andrzej Trautman on the occasion of his 64th Birthday, J.Tafel, editor.
  \MR{MR1691888 (2000c:83011)}

\bibitem{CCM2}
Y.~Choquet-Bruhat, P.T. Chru\'{s}ciel, and J.M. Mart\'in-Garc\'ia, \emph{{The
  Cauchy problem on a characteristic cone for the Einstein equations in
  arbitrary dimensions}}, Ann.\ H.\ Poincar\'e \textbf{12} (2011), 419--482,
  arXiv:1006.4467 [gr-qc]. \MR{2785136}

\bibitem{Ch-KL}
D.~Christodoulou and S.~Klainermann, \emph{On the global nonlinear stability of
  {M}inkowski space}, Princeton University Press, Princeton, 1995.

\bibitem{ChDelay}
P.T. Chru\'{s}ciel and E.~Delay, \emph{On mapping properties of the general
  relativistic constraints operator in weighted function spaces, with
  applications}, M\'em.\ Soc.\ Math.\ de France. \textbf{94} (2003), vi+103,
  arXiv:gr-qc/0301073v2. \MR{MR2031583 (2005f:83008)}

\bibitem{ChPaetz}
P.T. Chru\'{s}ciel and T.-T. Paetz, \emph{{The many ways of the characteristic
  Cauchy problem}}, Class.\ Quantum Grav. \textbf{29} (2012), 145006, 27,
  arXiv:1203.4534 [gr-qc]. \MR{2949552}

\bibitem{CorvinoSchoen2}
J.~Corvino and R.M. Schoen, \emph{On the asymptotics for the vacuum {E}instein
  constraint equations}, Jour.\ Diff.\ Geom. \textbf{73} (2006), 185--217,
  arXiv:gr-qc/0301071. \MR{MR2225517 (2007e:58044)}

\bibitem{Dossa97}
M.~Dossa, \emph{Espaces de {S}obolev non isotropes, \`a poids et probl\`emes de
  {C}auchy quasi-lin\'eaires sur un cono\"\i de caract\'eristique}, Ann. Inst.
  H. Poincar\'e Phys. Th\'eor. (1997), 37--107.

\bibitem{FRW}
H.~Friedrich, I.~R{\'a}cz, and R.M. Wald, \emph{On the rigidity theorem for
  space-times with a stationary event horizon or a compact {C}auchy horizon},
  Commun.\ Math.\ Phys. \textbf{204} (1999), 691--707, arXiv:gr-qc/9811021.

\bibitem{HIW}
S.~Hollands, A.~Ishibashi, and R.M. Wald, \emph{A higher dimensional stationary
  rotating black hole must be axisymmetric}, Commun.\ Math. Phys. \textbf{271}
  (2007), 699--722, arXiv:gr-qc/0605106.

\bibitem{MaertenKIDs}
D.~Maerten, \emph{Killing initial data revisited}, Jour.\ Math.\ Phys.
  \textbf{45} (2004), 2594--2599. \MR{MR2067575 (2005b:83009)}

\bibitem{RendallCIVP}
A.D. Rendall, \emph{Reduction of the characteristic initial value problem to
  the {C}auchy problem and its applications to the {E}instein equations},
  Proc.\ Roy.\ Soc.\ London A \textbf{427} (1990), 221--239. \MR{MR1032984
  (91a:83004)}

\bibitem{SemmelmannHab}
U.~Semmelmann, \emph{Conformal {Killing forms on Riemannian} manifolds},
  Habilitationsschrift,
  \url{www.igt.uni-stuttgart.de/LstGeo/Semmelmann/Publikationen/killing20.pdf}.

\bibitem{Semmelmann}
\bysame, \emph{Conformal {K}illing forms on {R}iemannian manifolds}, Math.\ Z.
  \textbf{245} (2003), 503--527. \MR{2021568 (2004i:53040)}

\bibitem{TilquinSchucker}
A.~Tilquin and T.~Schucker, \emph{{Maximal symmetry at the speed of light}},
  (2012), arXiv:1210.1468 [astro-ph.CO].

\end{thebibliography}
\end{document}